\documentclass[sigplan,10pt]{acmart}
\renewcommand\footnotetextcopyrightpermission[1]{}
\AtBeginDocument{%
  }

\usepackage{makecell}
\usepackage{caption}
\usepackage{subcaption}
\usepackage{algorithm}
\usepackage{algpseudocode}
\usepackage{booktabs}
\usepackage{array}
\usepackage{caption}
\usepackage{threeparttable} 
\usepackage{ragged2e}
\usepackage{fontsize} 
\usepackage{setspace} 
\usepackage{graphicx} 
\usepackage{subcaption} 
\usepackage{booktabs} 
\usepackage{enumitem}
\usepackage{xspace}
\usepackage{textcomp}
\usepackage{placeins} 
\usepackage{balance}
\AtEndDocument{\nobalance}

\usepackage{tikz}
\usetikzlibrary{positioning, arrows.meta}
\tikzset{
  >=Latex,
  box/.style={draw, rounded corners=2pt, fill=gray!8, inner xsep=6pt, inner ysep=5pt, align=center},
  link/.style={-Latex, line width=0.6pt}
}

\usepackage{tikz}
\DeclareRobustCommand{\fstep}[2][2.2ex]{%
  \tikz[baseline=-0.7ex]%
    \node[
      draw,circle,fill=black,
      minimum size=#1,      
      inner sep=0.6pt,
      text=white,
      font=\bfseries\footnotesize
    ]{#2};%
}

\usepackage{pifont}
\newcommand{\circnum}[1]{%
  \ifcase#1\or\ding{172}\or\ding{173}\or\ding{174}\or\ding{175}\or\ding{176}\or
  \ding{177}\or\ding{178}\or\ding{179}\or\ding{180}\or\ding{181}\fi} 

\newcommand{\ourSolution}{\textsc{SuperPass}\xspace}

\newcommand{\fastPathNoun}{fast track\xspace} 
\newcommand{\fastPathAdj}{fast-track\xspace}

\definecolor{darkyellow}{rgb}{0.8, 0.7, 0.0}  

\newcommand{\commentout}[1]{}

\begin{document}

\title{\LARGE{SuperPass: Fast-Tracking Blocking Threads\\ to Mitigate Priority Inversion on Mobile Devices}}


\author{Lei Li}
\affiliation{%
  \institution{City University of Hong Kong}
  \city{Hong Kong}
  \country{China}
}

\author{Yu Liang}
\affiliation{%
  \institution{Inria Paris}
  \city{Paris}
  \country{France}
}

\author{Riwei Pan}
\affiliation{%
  \institution{City University of Hong Kong}
  \city{Hong Kong}
  \country{China}
}

\author{Youcheng Sun}
\affiliation{%
  \institution{MBZUAI}
  \city{Abu Dhabi}
  \country{United Arab Emirates}
}

\author{Nan Guan}
\affiliation{%
  \institution{City University of Hong Kong}
  \city{Hong Kong}
  \country{China}
}

\author{Tei-Wei Kuo}
\affiliation{%
  \institution{Delta Electronics \& National Taiwan University}
  \city{Taipei}
  \country{Taiwan}
}

\author{Chun Jason Xue}
\affiliation{%
  \institution{MBZUAI}
  \city{Abu Dhabi}
  \country{United Arab Emirates}
}

\renewcommand{\shortauthors}{Trovato et al.}

\begin{abstract}
Priority inversion occurs when a high-priority thread is delayed by a lower-priority one.
Although well studied in real-time systems, its impact in general-purpose OSes (e.g., Android) remains underexplored.
On Android, we find that priority inversions happen frequently and can delay latency-critical threads, degrading user experience. For example, the foreground app’s UI thread is frequently blocked by low-priority threads, with blocking durations of up to 210\,ms, enough to cause dropped frames.
Existing solutions designed for real-time systems fail to eliminate long priority-inversion blockings on latency-critical threads and may introduce high overhead on Android.

To solve this problem, we uncover two insights on Android: 1) long blockings are mainly due to the accumulated CPU waiting time of low-priority blocking threads rather than their critical-section latency; and 2) although latency-critical threads can be blocked by many concurrent readers, tracking a limited number of them is sufficient to achieve good responsiveness with low overhead in most cases.
Guided by these insights, we propose \ourSolution, a lightweight kernel mechanism that mitigates priority inversion by \fastPathAdj scheduling of low-priority threads blocking latency-critical threads.
It introduces a scheduler \fastPathNoun that grants immediate CPU access to threads blocking latency-critical threads, and employs a lock-level detector that effectively identifies most such blocking threads.

We evaluate \ourSolution on a Google Pixel~8 smartphone. Taking UI thread as a case study, \ourSolution decreases the 99.9th-percentile blocking duration by 72.0\% and blocking count by 47.7\% on average compared to the default scheduler, and reduces janky frames by 29.2\% with a system-wide CPU overhead of only 0.74\%. \ourSolution also outperforms existing approaches including priority inheritance, real-time UI promotion, and Proxy Execution.

\end{abstract}

\settopmatter{printfolios=true}
\maketitle
\pagestyle{plain}

\section{Introduction}
The priority inversion problem occurs when a higher-priority thread is blocked by a lower-priority thread for a potentially unbounded period~\cite{PIP_PCP_1990,priorityInversionFormation,PI_concept1,PI_concept2, liu2020removing}. This issue  typically arises when using synchronization primitives such as mutexes, seqlocks, and semaphores. It is a critical and intolerable problem in real-time systems because these systems must meet specific deadlines~\cite{rt-important,linux-not-important}. However, the prevalence and severity of priority inversion problems in general-purpose OSes such as Android remain largely unknown.

\begin{figure}[t]
\vspace{0.5em}
  \centering
  \begin{subfigure}[t]{0.47\textwidth}
    \centering
    \includegraphics[width=\textwidth]{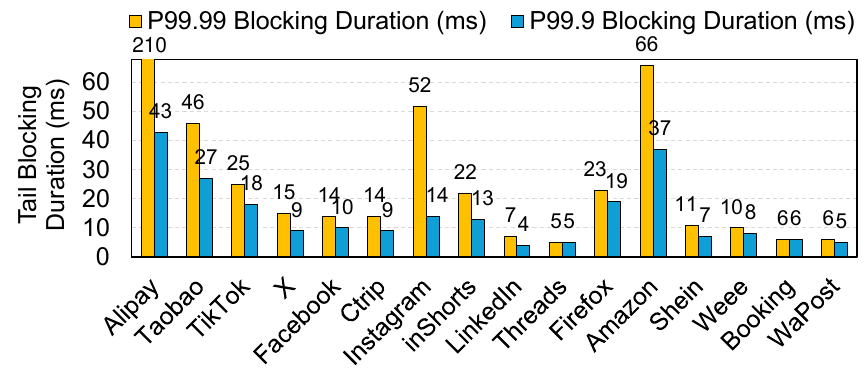}
    \caption{
    The 99.9th and 99.99th percentile blocking durations caused by priority-inversion on UI thread.}
    \label{fig:motiv_tail}
  \end{subfigure}
  
  
  \begin{subfigure}[t]{0.47\textwidth}
    \centering
    \includegraphics[width=\textwidth]{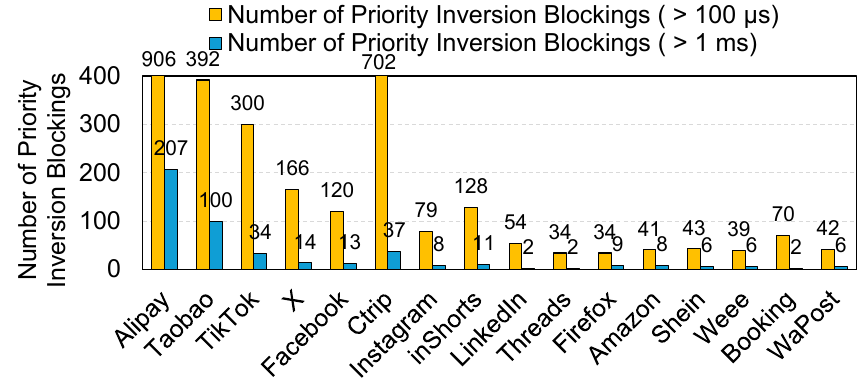}
    \caption{The number of priority-inversion blockings exceeding 100\,\textmu s and 1\,ms on  UI thread (one-minute testing intervals).}
    \label{fig:motiv_count}
  \end{subfigure}

  \vspace{-0.5em}
  
  \caption{Priority inversion blockings on latency-critical threads (foreground UI thread as a case study). Priority inversions can last long enough to cause frame drops.
  }
  \label{fig:motiv}
  \vspace{-1.5em}
\end{figure}

\noindent\textbf{Priority inversion is prevalent and severe on Android.} 
To fill this gap, we first quantify priority inversions on multiple Android smartphones. Our measurements show that the priority inversion problem on latency-critical threads occurs commonly. As the number and complexity of applications continue to grow, the problem becomes more severe. This indicates that mitigating priority inversion in general-purpose OSes (e.g., Android) is necessary and urgent.
In this paper, we use the foreground app's UI thread as a case study of latency-critical threads. 
Its prolonged latency, missing the frame-rending deadline, could induce user-visible frame drops and negatively affect user experience~\cite{UIjank1, UIjank2}.
Figure~\ref{fig:motiv} shows that priority inversions on UI thread occur frequently and can last up to 210\,ms.
On a 120\,Hz display, a 210\,ms blocking interval corresponds to at least 20\% frame loss within one second.
The problem also appears in other latency-critical threads (e.g., input dispatch, sensor processing, camera pipeline control, and XR/AR tracking) because these threads can also be blocked by low-priority threads.

\noindent\textbf{Existing solutions are ineffective on Android.} 
Existing solutions for priority inversion in real-time systems include priority ceiling protocol (PCP) and priority inheritance (PI)~\cite{PIP_PCP_1990,goodenough1988priority,cheng2007implementation,PIP2,PCP2,multiprocessor2SharedMemoryMultiprocessors,multiprocessor4RealTimeTasks}. PCP requires prior knowledge of all tasks and their priorities, which is impractical for dynamic Android workloads. PI temporarily boosts a lock holder's priority to that of the blocked thread. However, we find that existing real-time solutions fail to eliminate long priority-inversion blockings on latency-critical threads and can incur high overhead on Android for two reasons. First, Android's scheduler does not guarantee immediate CPU access after raising priority, so a boosted blocking thread may still wait behind other runnable threads. Second, PI is typically applied to exclusive locks with a single holder, whereas latency-critical threads on Android are often blocked by multiple concurrent holders, making such boosting both ineffective and costly.

\noindent\textbf{Solving priority inversion on Android is non-trivial.} 
Two challenges must be overcome. \emph{Accelerating blocking threads effectively}: priority boosting alone is insufficient, and directly manipulating the scheduler's ordering key risks interfering with real-time threads and starving unrelated work on resource-constrained mobile devices. \emph{Identifying blocking threads efficiently}: a single priority inversion can span multiple locks and threads through nested dependencies, and Android's diverse lock types and vendor-specific call paths make per-lock instrumentation impractical. Moreover, a latency-critical thread may be blocked by up to 24 concurrent readers, making full dependency tracking online prohibitively expensive on mobile devices.

\noindent\textbf{Key insights.} 
We uncover two insights that help navigate these challenges. 
\emph{Insight 1: runnable delays dominate priority-inversion blockings.} Long blockings are mainly due to accumulated CPU waiting time of low-priority blocking threads in the scheduler runqueue rather than their critical-section latency--at P99.9, the gap between runnable delay and critical-section latency reaches 1000X. This explains why PI-based solutions are ineffective, as raising a thread’s priority has limited impact on vruntime, the thread may still need to remain in the runqueue for a substantial amount of time.
\emph{Insight 2: tracking a limited number of readers is sufficient.} Although latency-critical threads can be blocked by many concurrent readers, tracking a limited number of them is sufficient to achieve good responsiveness with low overhead in most cases, since 98\% of inversions involve fewer than 8 concurrent readers in practice.

\noindent\textbf{\ourSolution.} 
Guided by these insights, we propose \ourSolution, a lightweight kernel mechanism that mitigates priority inversion by \fastPathAdj scheduling of low-priority threads blocking latency-critical threads. \ourSolution comprises two co-designed components: a scheduler \fastPathNoun and a lock-level detector. The \fastPathNoun grants immediate CPU access to blocking threads by manipulating the scheduler's ordering key, bypassing normal scheduling competition without interfering with real-time classes, and uses lightweight safeguards that bound the scope and duration of acceleration to limit impact on fairness. The lock-level detector efficiently identifies most blocking threads without reconstructing full dependency graphs, covering both exclusive and shared locks, alongside an offline lock-identification module that eliminates per-lock instrumentation by recording lightweight sleep fingerprints at a single scheduler hook point.

\noindent\textbf{Results.}
We evaluate \ourSolution on a Google Pixel~8 smartphone using a variety of commonly-used applications. 
Taking the UI thread as a case study, the results show that \ourSolution decreases the 99.9th-percentile blocking duration by 72.0\% and blocking count by 47.7\% on average compared to the default scheduler, and reduces janky frames by 29.2\% with a system-wide CPU overhead of only 0.74\%. \ourSolution also outperforms existing approaches, including priority inheritance, real-time UI promotion, and Proxy Execution.

\noindent\textbf{Contributions.} 
The contributions of this paper are as follows: 

\begin{itemize}
\vspace{-0.7em}
  \item We evaluate the prevalence and impact of long priority-inversion blockings on latency-critical threads on Android smartphones (\S~\ref{motivation});
\vspace{-0.5em}  
  \item We identify why existing solutions fail on general-purpose OSes (\S~\ref{motivation}), and uncover two insights to enable an efficient solution (\S~\ref{insights});
\vspace{-0.5em}   
  \item We propose \ourSolution, a low-overhead mechanism that efficiently identifies low-priority threads that blocks latency-critical threads and accelerates their execution through a novel \fastPathAdj scheduling strategy (\S~\ref{design});
  \vspace{-0.5em}  
  \item We implement and evaluate \ourSolution on a Google Pixel~8 (Linux~5.15, Android~14)
  and observe significant reductions in blocking duration and dropped frames in UI-thread instantiation compared to state-of-the-art solutions (\S~\ref{evaluation}).

\end{itemize}

\section{Background and Motivation} \label{motivation}

To study priority inversion on Android, we use the foreground UI thread as a representative latency-critical thread. 
It initiates frame production (input callbacks, layout, and draw command recording) under a per-frame deadline set by the display refresh rate (e.g., 8.3\,ms at 120\,Hz). Blocking on this thread can miss the deadline and cause \emph{janky frames} (dropped or delayed frames with visible stutter)~\cite{UIjank1,UIjank2}.


\subsection{Prevalence and Impact of Priority Inversions on Android} 
\label{longUIblock}
We take the foreground UI thread as a case study to investigate the prevalence and impact of priority inversion on Android. 
We evaluate the UI threads of 16 popular applications on a Google Pixel 8. In each evaluation, we interact with one foreground app for one-minute interval, while 12 apps run concurrently in the background.
We repeat this process ten times per app and automate the interactions using adb shell input commands.

Figure~\ref{fig:motiv} shows that priority inversions occur frequently on UI threads across popular apps. Figure~\ref{fig:motiv_tail} presents the 99.9th/99.99th percentile blocking durations with tail latencies up to \(\sim\)210\,ms; Figure~\ref{fig:motiv_count} shows 
that each app experiences priority inversions, with many exhibiting long tails well beyond the 8.3\,ms frame budget at 120\,Hz, implying dropped frames (Google Pixel 8 supports refresh rates up to 120\,Hz~\cite{pixel8wiki}). For example, Alipay reaches P99.99 \(\approx\)210\,ms with \(\sim\)906 inversions exceeding 100\,\text{\textmu s} and 207 inversion blockings exceeding 1\,\text{ms} per minute. Across apps, \(>100\,\text{\textmu s}\) events occur in dozens to hundreds per run, and \(>1\,\text{ms}\) events often exceed 30. Overall, priority inversions are common and long enough to cause frame janks.

\begin{figure}[!h]
\vspace{-1em}
\centering
\includegraphics[width=0.48\textwidth]{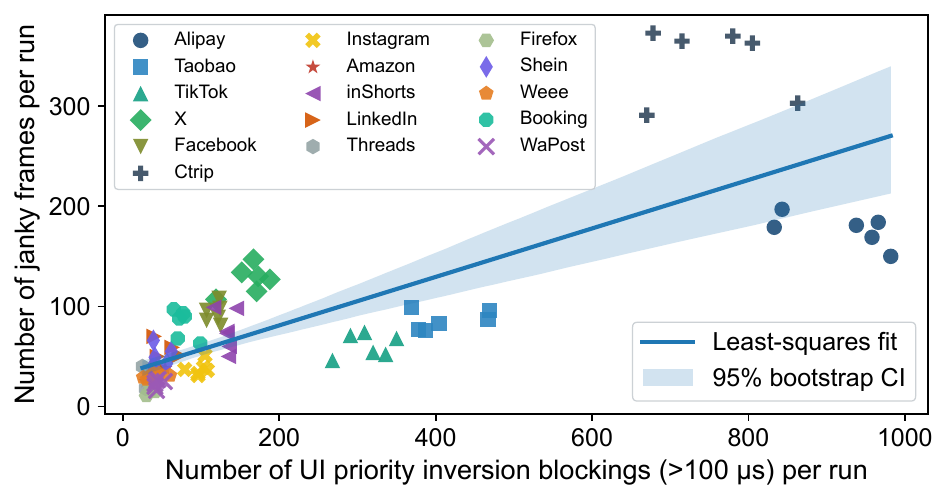}
\vspace{-2.3em}
\caption{Priority inversions correlate with janky frames. Each point is one run.
Points with the same marker and color correspond to the same app. The solid line is a least-squares fit with a 95\% bootstrap confidence interval (shaded). 
Across runs, the correlation is strong (Pearson $r=0.79$, Spearman $\rho=0.84$).} 
\vspace{-1.5em}
\label{fig:motiv_correlation}
\end{figure}

Janky frames directly degrade user-perceived UI smoothness and responsiveness. We quantify how UI-thread priority inversions translate into janky frames under the same setup as Figure~\ref{fig:motiv}. For each one-minute run, we measure (i) the number of priority-inversion blockings on UI threads with blocking times exceeding $100\,\text{\textmu s}$, and (ii) the number of janky frames that overlap with these blocking intervals (i.e., janky frames directly extended by priority inversions on UI thread ). We run each app ten times and randomly show 6 runs per app for clarity. 
Figure~\ref{fig:motiv_correlation} shows a clear relationship between them: runs with more UI-thread priority inversions tend to exhibit more janky frames, consistent across apps despite varying per-app correlation. Across runs, the correlation is strong (Pearson $r=0.79$, Spearman $\rho=0.84$).

\subsection{Limitations of Existing Solutions}
\label{intuitiveSolution}
\noindent\textbf{Existing Solutions for Priority Inversions.} In real-time systems, there are two types of classic existing solutions for priority inversion: priority inheritance (PI) and priority ceiling protocol (PCP)~\cite{PIP_PCP_1990,goodenough1988priority,cheng2007implementation}. 
PI temporarily raises a low-priority lock holder to the blocked thread’s priority until release, and to propagate this boost transitively across chains of blocking threads. PCP assigns each lock a ceiling equal to the highest priority of any task that may use it; a task that acquires the lock runs at this ceiling until release.
PCP requires prior knowledge of all potential lock users and their priorities, which is impractical in dynamic Android workloads. Therefore, our evaluation focuses on PI. We optimize PI for the Linux read/write semaphore (rwsem), boosting only the priority of the writer holding the lock (writer holder). 

We also evaluate an intuitive baseline that treats the UI thread as a real-time thread (RT-UI). 
Android divides threads into two classes: real-time threads (priority 0--99) and normal threads (priority 100--139).
Real-time threads are scheduled by the real-time scheduler, while normal threads are scheduled by CFS~\cite{CFSone, CFStwo}. 
Unlike real-time schedulers that always select the highest-priority task, CPU scheduler on Android (i.e., CFS) schedules threads based on virtual runtime (\emph{vruntime}) rather than only priority to balance priority and fairness. RT-UI moves the UI thread from the normal (CFS) class into the real-time class, allowing it to preempt normal threads whenever it is runnable.

All experiments in this section follow the same setup and procedure described in \S~\ref{longUIblock}.
We first present experimental results that expose the limitations of PI and RT-UI, and then explain the two root causes of the limitations on Android.

\begin{figure}[h!]
\centering
\vspace{-1em}
\includegraphics[width=0.47\textwidth]{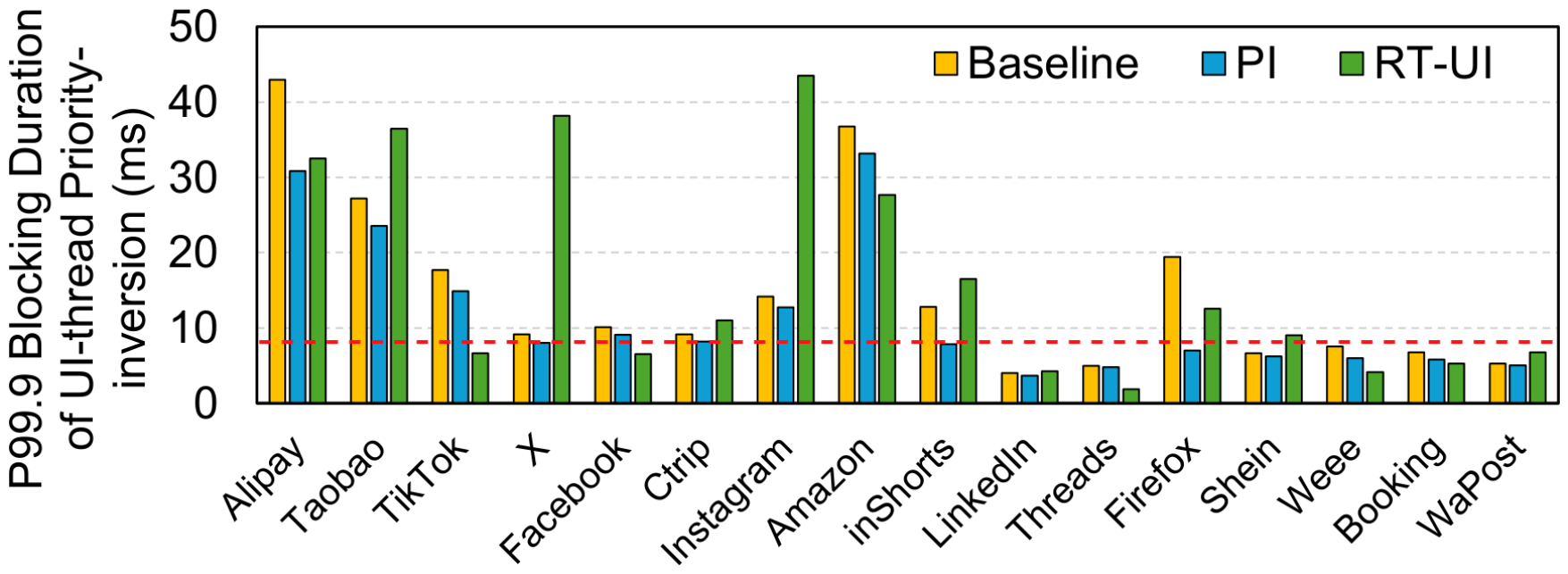}
\vspace{-1em}
\caption{
P99.9 blocking duration of priority inversions on UI threads across 16 apps under Baseline, PI, and RT-UI. 
The red horizontal line at 8.3\,ms marks the per-frame budget at 120\,Hz. A block duration exceeding this latency will miss the 120\,Hz deadline and result in a dropped frame (i.e., jank).}
\label{fig:motiv_otherSolution}
\vspace{-1em}
\end{figure}

\noindent\textbf{Limitations and Penalties of Existing Solutions.} Figure~\ref{fig:motiv_otherSolution} shows P99.9 blocking duration of priority inversions on UI threads across 16 apps (for brevity, we show only P99.9 here; full results appear in \S~\ref{evaluation}). 
The results show that although PI can reduce the P99.9 blocking duration by an average of 15.6\% across 16 apps, it leaves P99.9 blocking durations on some apps over 8.3 ms, causing janky frames. 
PI is ineffective on Android for two reasons.
First, raising the priority of the low-priority blocking threads did not make UI thread to run on CPU instantly.
Second, to reduce overhead, PI is usually used for exclusive locks to raise the priority of one lock holder. However, latency-critical threads on Android are usually blocked by shared locks with multiple lock holders. 

While RT-UI benefits some apps, it degrades others.
RT-UI promotes the UI thread to the real-time class, but it does not accelerate the low-priority blocking threads of UI thread, so the UI thread can remain blocked.
It can even extend the priority-inversion blocking duration because it increases real-time scheduling pressure, which leaves less CPU time for CFS-scheduled blocking threads of UI thread to run and release the lock.
Thus, existing approaches fail to consistently eliminate long priority-inversion blockings on UI threads. 

\begin{figure}[h!]
\centering
\vspace{-0.6em}
\includegraphics[width=0.46\textwidth]{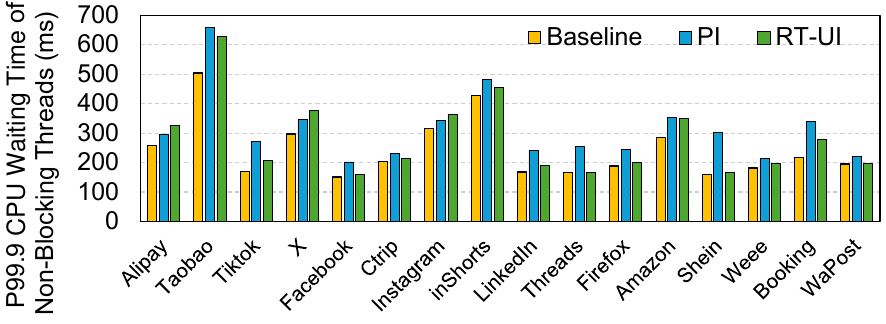}
\vspace{-1.2em}
\caption{P99.9 CPU waiting time for non-blocking low-priority threads across 16 apps under Baseline, PI, and RT-UI.}
\vspace{-0.8em}
\label{fig:PI_longer_runnable}
\end{figure}

To show existing solutions' impact on low-priority threads that do \emph{not} block the UI thread (i.e., non-blocking threads), we measure the P99.9 CPU waiting time (i.e., time to be scheduled on the CPU) of non-blocking threads over each one-minute run.
Figure~\ref{fig:PI_longer_runnable} shows that PI increases this delay by 8.8\%--88.1\% (mean 32.6\%) versus Baseline, while RT-UI increases it by 0.6\%--28.4\% (mean 13.7\%). 
This means that both PI and RT-UI increase the waiting time of non-blocking threads for CPU access.
PI raises the priority of low-priority threads whenever a priority inversion is detected. 
As a result, many low-priority threads with raised priority compete for CPU time, which increases CPU waiting time of non-blocking threads.
For RT-UI, promoting the UI thread to the real-time class allows it preempt normal threads more often, leaving less CPU time for normal-class non-blocking threads.

In summary, existing approaches fail to consistently eliminate long priority-inversion blockings on UI threads and may increase the CPU waiting time of non-blocking threads, thereby impacting fairness.

\section{Insights and Analysis} \label{insights}

In this section, we first breakdown the long priority-inversion blocking duration of latency-critical threads to pinpoint the dominant contributor. 
We then analyze the shared lock, which is commonly used on Android to understand the difficulty of tracking concurrent reader holders online. Based on this analysis, we uncover two insights that directly guide the design of \ourSolution.

\subsection{Accumulated Runnable Delays Dominate Priority Inversion Blockings}

To solve priority inversion problems, we break down the blocking duration of latency-critical threads to identify the dominant components. 

\begin{figure}[h!]
\centering
\vspace{-0.8em}
\includegraphics[width=0.47\textwidth]{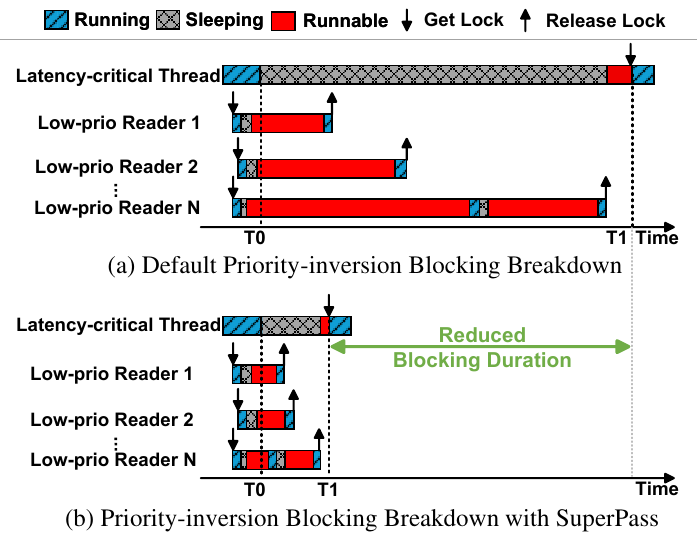}
\vspace{-0.8em}
\caption{Priority-inversion blocking breakdown. 
\ourSolution reduces the runnable delay of low-priority reader holders, and thus shortens the priority-inversion blocking duration of the latency-critical thread.}

\vspace{-1.3em}
\label{fig:design_long_runnable_problem}
\end{figure}

\noindent\textbf{Blocking Breakdown.}
Figure~\ref{fig:design_long_runnable_problem}(a) illustrates a representative priority-inversion blocking breakdown involving a contended reader-writer lock (\texttt{rwsem}).
At time $T0$, a latency-critical thread (top row) requests the lock and begins waiting.
The lock is held by a set of low-priority reader holders and may also be delayed by queued waiters (e.g., a writer). 
The latency-critical thread can acquire the lock only after all current reader holders release it and any queued waiters ahead of it have completed and released the lock, at T1.
For each blocking thread on the critical dependency chain, the time to reach lock release consists of (i) its on-CPU execution (running), (ii) time spent sleeping (e.g., waiting for I/O or another lock), and (iii) \emph{runnable delay} after wakeups. We call a thread's waiting time in CPU scheduler runqueue a \emph{runnable delay}: the time from when a thread is enqueued on the runqueue to when it first runs on a CPU core. 
Figure~5(a) shows concurrent readers running on different CPU cores, so the latency-critical thread waits for the slowest reader. In the worst case, if multiple readers are serialized on one core, their delays can accumulate and further extend the blocking duration.

\begin{figure}[h!]
\centering
\includegraphics[width=0.47\textwidth]{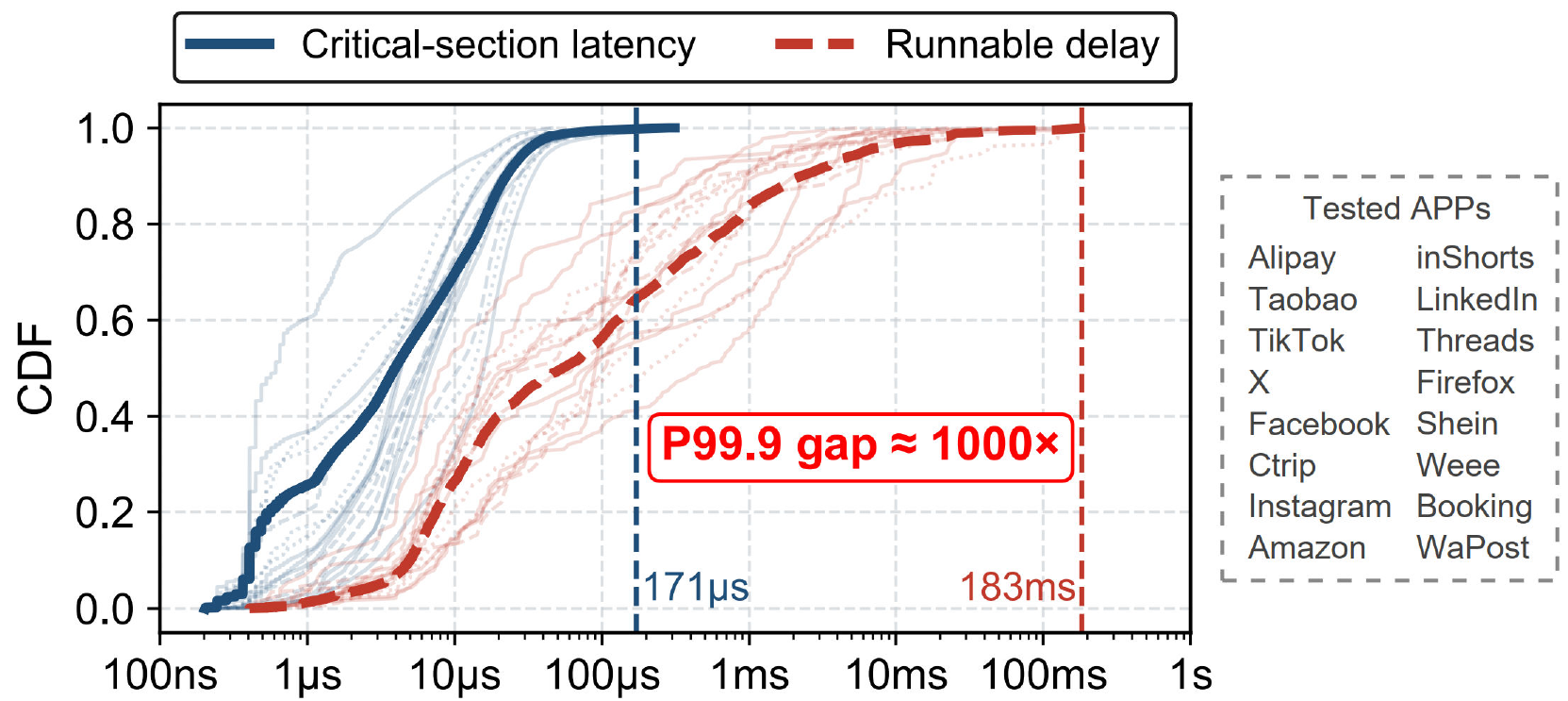}
 \vspace{-0.9em}
\caption{Critical-section latency vs. runnable delay across 16 apps. The x-axis uses a logarithmic scale to cover latencies from nanoseconds to seconds. The bold curves are calculated from all results merged across the 16 apps, while the light curves show the per-trace results for individual app. 
At P99.9, runnable delay is $1000\times$ of critical-section latency.}
\vspace{-2em}
\label{fig:critical_section_vs_runnable}
\end{figure}

\noindent\textbf{Quantifying runnable-delay accumulation and critical-section latency.}
Profiling 16 popular apps under a realistic load (one foreground app with 12 background apps for one minute), we quantify both runnable delays and the lock-holding critical-section latency, and show how runnable delays accumulate along the critical path.
First,  a \textbf{single} runnable delay often reaches the millisecond scale, frequently exceeds $50$\,ms, and has P99.9 up to $183$\,ms (Figure~\ref{fig:critical_section_vs_runnable}).
In contrast, the corresponding critical-section latencies are typically in the microsecond-to-millisecond range, with P99.9 up to $171$\,\textmu s, yielding up to a $1000\times$ gap (Figure~\ref{fig:critical_section_vs_runnable}). 
At P90 the gap is around $100\times$.

Second, one blocking thread can incur \textbf{multiple} runnable delays within one priority inversion because it may repeatedly sleep (e.g., waiting for I/O or another lock) and wake before releasing the lock; each wakeup re-enqueues the thread and may incur a runnable delay. We observe that $13.2\%$ of blocking threads have such repeated sleep-wake behavior (see supplementary file), amplifying the accumulation of runnable delay.
Third, one priority inversion often involves \textbf{multiple} blocking threads (e.g., multiple concurrent reader lock holders and queued lock waiters), further compounding runnable delays on the latency-critical threads’ critical path (Figures~\ref{fig:analysis_reader_owner_cdf}).
Based on these results, we uncover a new insight.

\noindent\textbf{Insight~1: \emph{Runnable delays of blocking threads induced by general-purpose OSes' (e.g., Android) schedulers dominate the priority inversion blocking duration of latency-critical threads.}}

This insight helps explain why priority inheritance (PI)-based solutions are ineffective on Android. They raise the priority of a low-priority thread but do not guarantee to decrease its runnable delays.
PI assumes priority-driven scheduling, where elevating priority instantly translate into CPU access, but Android’s CFS schedules threads by \emph{vruntime} for fairness.
Thus, even after PI raises a blocking thread’s priority, it may still wait behind other runnable threads, resulting in long blocking durations for latency-critical threads.

\subsection{Tracking Reader Holders is Feasible on Android}

\begin{figure}[!h]
\centering
\includegraphics[width=0.41\textwidth]{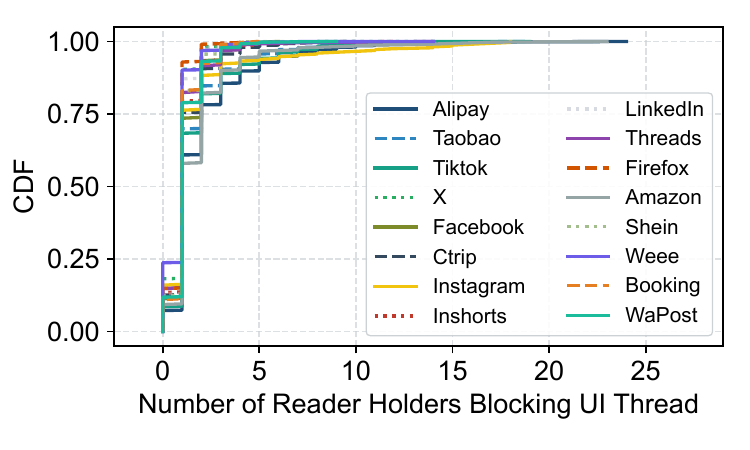}
\vspace{-1.8em}
\caption{Concurrent reader holders blocking UI thread are usually small (\(\approx 98\%\) have $\leq$8.)}
\vspace{-1.5em}
\label{fig:analysis_reader_owner_cdf}
\end{figure}

To understand the complexity and overhead of identifying \emph{all} reader holders online, we analyze how many concurrent readers a thread typically has in practice.
Based on profiling results from a Google Pixel 8 running multiple commonly used apps, we make two observations.
First, Android apps heavily use shared locks such as \texttt{rwsem}, where latency-critical threads could be blocked by many concurrent reader holders, up to 24, as shown in Figure~\ref{fig:analysis_reader_owner_cdf}. 
Second, in most cases, the number of concurrent reader holders is small, with 98\% having fewer than 8. 
This likely reflects the limited hardware parallelism and practical concurrency of smartphone workloads (e.g., our tested smartphone has 9 CPU cores).

In real-time systems, meeting strict deadlines often requires identifying and accelerating \emph{all} reader holders that can block a latency-critical thread. Our experiments on Android show in Figure~\ref{fig:analysis_reader_owner_cdf}, where we found up to 24 concurrent readers. Identifying all reader holders can incur substantial overhead and complicate worst-case blocking bounds and schedulability analysis.
However, Android’s latency-critical threads target responsiveness rather than strict hard-real-time guaranties, so it is practical to track and accelerate only a few readers that cover most cases to reduce overhead.


\noindent\textbf{Insight~2: \emph{On Android, although latency-critical threads could be blocked by many readers, tracking a limited number of readers suffices to achieve good responsiveness while maintaining low overhead in most cases.}}

This insight is mostly applicable to mobile devices, which have a relatively low number of CPU cores and limited hardware parallelism. 
On platforms with many more CPU cores and higher concurrency, a latency-critical thread may be blocked by a much larger set of concurrent readers across multi-layer dependency chains, making online tracking more costly and bounded tracking less effective.

\section{Design of \ourSolution} \label{design} 


As discussed above, priority inversion can block latency-critical threads (e.g., the UI, input dispatch, and camera pipeline control threads) and thus severely hurt responsiveness on Android. This paper proposes \ourSolution, a lightweight kernel mechanism that mitigates such blocking by identifying low-priority blocking threads and accelerating their execution, allowing them to release locks earlier.

\begin{figure}[h!]
\centering
\includegraphics[width=0.485\textwidth]{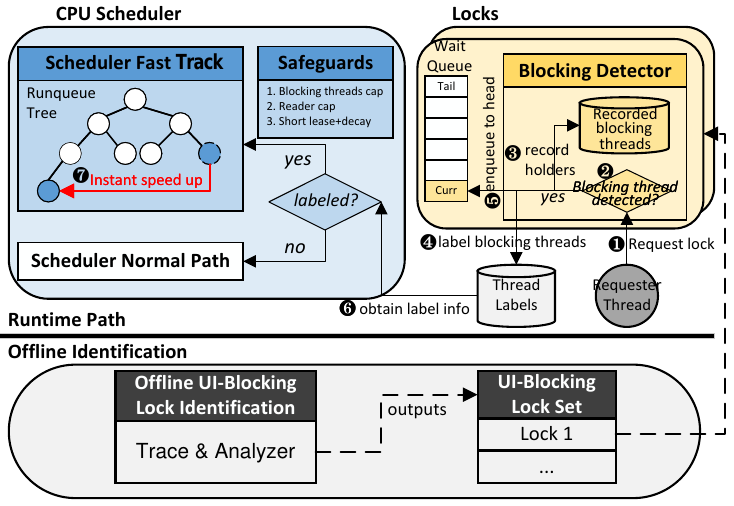}
\vspace{-0.5em}
\vspace{-1.5em}
\caption{Design overview of \ourSolution.}

\label{fig:design_overview}
\vspace{-2em}
\end{figure}

\noindent\textbf{Overview.}
Figure~\ref{fig:design_overview} presents the design overview of \ourSolution, which comprises four components:
(i) a scheduler \fastPathNoun (\S~\ref{fastPathOnly}) that instantly accelerates the execution of low-priority threads blocking the latency-critical threads (blue panel in Figure~\ref{fig:design_overview}); 
(ii) scheduler safeguards (\S~\ref{sec:safeguards}) that bound the acceleration and limit impact on non-blocking threads (blue panel); 
(iii) a lock-level detector (\S~\ref{blockerDetector}) that identifies blocking threads of the latency-critical threads (yellow panel); and 
(iv) a portable lock-identification module (\S~\ref{lockPinpoint}) that identifies the locks where the latency-critical threads block on (gray panel).


\subsection{Fast-Track Scheduling for Blocking Threads}\label{fastPath}
Guided by \textbf{Insight~1}, \ourSolution focuses on the dominant contributor to long priority-inversion blockings: scheduler induced runnable delays of blocking threads. \ourSolution therefore accelerates blocking threads in the scheduler to cut their runnable delays using a scheduler \fastPathNoun. 
However, this path can increase contention and delay unrelated work. Thus, we add lightweight safeguards to limit how many low-priority threads are expedited and for how long.

\subsubsection{Fast Track: Accelerating Blocking Threads}\label{fastPathOnly}

A naive solution to long blockings on latency-critical threads is to make all blocking threads real-time, but this risks fairness for other real-time tasks, especially when many blocking threads are present. Instead, we design a \fastPathNoun mechanism that allows designated low-priority blocking threads to bypass normal scheduling competition and run immediately, without interfering with real-time classes. As illustrated in Figure~\ref{fig:design_long_runnable_problem} (b), accelerating the blocking thread reduces its runnable delays (red rectangles) and thus shortens the priority-inversion blockings of the latency-critical threads.

Android CFS schedules threads based on virtual runtime (\emph{vruntime}) rather than priority. \emph{vruntime} represents the amount of CPU time a thread has consumed relative to its weight. 
CFS tries to allocate each thread the same amount of virtual runtime to enforce fairness, so it always selects the thread with the smallest \emph{vruntime}.
In this case, a thread with raised priority may still wait in the runqueue for a long time if its \emph{vruntime} remains larger than others.

To address this issue, SuperPass moves labeled blocking threads to the head of the runqueue order so they run sooner. We will introduce how \ourSolution identifies and labels blocking threads in \S\ref{ApproxBlockTree}. Specifically, \ourSolution adds a per-thread flag in the kernel task structure. Only threads with this flag use the \fastPathNoun, and all other threads follow normal scheduling. The \fastPathNoun operates as follows. When a labeled blocking thread  is enqueued to the scheduler runqueue (Step~\fstep{6} in Figure~\ref{fig:design_overview}), its scheduling metric (e.g., \emph{vruntime} in CFS) is reset to place it at the head of the runqueue for immediate scheduling (Step~\fstep{7}). Once the priority inversion ends, the label is cleared and its original state is restored. In addition, the labeled blocking thread is protected from involuntary preemption (e.g., by suppressing timeslice expiration), ensuring it can complete its critical section promptly. 
These changes reduce the runnable delay of threads blocking latency-critical threads, allowing them to run on the CPU earlier and release locks sooner.

\subsubsection{Safeguarding Fairness}\label{sec:safeguards}
Blindly accelerating many blocking threads at once (e.g., concurrent reader holders and queued waiters) can increase CPU competition among them and significantly delay other unrelated threads.
To reduce the interference on those unrelated threads, previous approaches employ complicated and high overhead reservation frameworks~\cite{abeni1998cbs,schedDeadline,hierarchicalServers}.
However, the overhead is too large for mobile devices, thus we use three lightweight safeguards to bound who is expedited and for how long: 

(i) \textit{Limit the scope of the acceleration to the latency-critical threads' blocking threads only.} Unlike PI, which blindly boosts the priority of any low-priority lock holder that blocks a higher-priority task, \ourSolution accelerates only the low-priority threads blocking the latency-critical threads. Threads not blocking the latency-critical threads are never accelerated to limit the impact on unrelated threads;  

(ii) \textit{Cap on \fastPathAdj threads.} Limit the number of \fastPathAdj blocking threads to one fewer than the available online CPU cores, ensuring at least one core remains available for non-blocking threads while avoiding CPU competition among \fastPathAdj blocking threads; 


(iii) \textit{Deadline-aligned acceleration lease.} 
Revoke labels at least once per latency budget interval \(T_{\text{budget}}\), because acceleration is intended to shorten the latency-critical threads' blocking duration, rather than to keep blocking threads continuously accelerated.
On revocation, restore the thread’s original scheduling metric and apply a short cool-down period before re-labeling to avoid immediate repeated acceleration of the same thread. 

These safeguards bound the scope and duration of \fastPathNoun acceleration, preserving system-wide responsiveness while retaining the gains from \fastPathAdj scheduling.

\noindent\textbf{Implementation.}
For UI-thread instantiations, \(T_{\text{budget}}\) is set to the frame interval (e.g., 16.7\,ms at 60\,Hz and 8.3\,ms at 120\,Hz). For other latency-critical threads, it can be set to the corresponding service deadline or latency target.
We set the cool-down period to 1\,ms for the UI-thread instantiation as a small fraction of the frame budget (8.3--16.7\,ms). This is long enough to prevent immediate repeated re-acceleration of the same blocking thread, but short enough to allow re-labeling within the same frame if the thread blocks the UI thread again.

\subsection{Identifying Blocking Threads and Locks}\label{ApproxBlockTree}
In \S~\ref{fastPath}, we showed how \ourSolution reduces blocking duration on latency-critical threads once the right blocking threads are identified. 
This section describes how we identify blocking threads and the locks on which they are blocked.

\subsubsection{Lightweight Identification of Blocking Threads} \label{blockerDetector}

A single priority inversion can span multiple locks and threads, so reconstructing full dependencies online is expensive.

However, guided by \textbf{Insight~2}, we avoid full dependency tracking on Android to identify lock holders with low overhead. 
Concretely, our lock-level scheme (i) labels only a bounded set of low-priority \emph{holders}, and (ii) bypasses lock \emph{waiters}. We classify participants at a contended lock into holders (writer/reader holders) and waiters, and handle them differently because holders already own the lock, while waiters have not acquired it yet. 
We first present the detection and labeling workflow, and then describe how we handle different lock participants (holders vs.\ waiters) under contention.

\noindent\textbf{Detection and Labeling Workflow.} 
Algorithm~\ref{alg:identifyBlockTree} illustrates how we deal with holders and waiters. The detector runs only when a lock acquisition fails (Line 3; Step \fstep{1} in Figure~\ref{fig:design_overview}). It first checks whether the requester is the latency-critical thread or if it is labeled before (Line 4), then verifies that the current holder has lower priority (Line 5; Step \fstep{2}). If both conditions hold, the holder is recorded (Step \fstep{3}) and labeled (Step \fstep{4}) to use the scheduler \fastPathNoun (Line 6). Waiters are bypassed by placing the request thread at the head of the wait queue (Lines 9-10; Step \fstep{5}). When the blocked requester eventually acquires the lock, the label is cleared (Lines 11-12). 
In Line 5, we also check the "already labeled" thread because the “already labeled” predicate propagates criticality transitively along dependency chains (i.e., if a thread is labeled because it blocks the latency-critical threads or a previously labeled thread, then any holder that now blocks this requester is  blocking thread of the latency-critical threads.) This gives us multi-hop coverage (nested locks, subsystem calls) without building a full dependency graph.

\begin{algorithm}[h!]

\small
\caption{Lock-Level Labeling of Blocking Threads}
\label{alg:identifyBlockTree}
\begin{algorithmic}[1]
\State \textbf{Input:} \textit{Curr} = thread requesting the lock;
\State \textbf{Init:} \textit{lock\_holder}; \Comment{set to record labeled lock holder}

\State On \textit{Curr} lock request failed:
\If{\textit{Curr} is  latency-critical thread or already labeled}
    \If{\textit{Curr} has lower priority than \textit{lock\_holder}}
        \State record \textit{lock\_holder} and label it;
    \EndIf
\EndIf

\State On \textit{Curr} start waiting:
\State \hspace{\algorithmicindent} Place \textit{Curr} at the head of the lock’s wait queue;

\State On \textit{Curr} lock request succeed: 
\State \hspace{\algorithmicindent} Clear the label of the recorded \textit{lock\_holder};

\end{algorithmic}
\end{algorithm}


\noindent\textbf{Lock Holders.} 
Identifying lock holders is straightforward for mutexes because there is a single owner, which is stored in a variable (each thread getting a lock will set the variable, and clear the variable when it releases the lock). 
However, for shared locks like \texttt{rwsem}, multiple readers may hold the lock concurrently, as shown in Figure~\ref{fig:analysis_reader_owner_cdf}. 
In real-time systems, accelerating all concurrent readers is needed to meet deadlines.
However, tracking \emph{all} readers by inserting/removing an owner record on every acquire/release incurs substantial CPU and memory overhead~\cite{linux_rwsem_owner_overhead}. 
In contrast, Android’s latency-critical threads target responsiveness rather than hard real-time guarantee.
Guided by \textbf{Insight~2}, we therefore track only a limited set of reader holders with low overhead, up to $N_{\text{CPU}}-1$), which covers most cases.
All recorded holders are labeled and accelerated via the scheduler \fastPathNoun.

\noindent\textbf{Lock Waiters.} 
Lock holders already own the lock and are executing the critical section, so the blocked latency-critical thread must wait for them to release it and cannot bypass them. In contrast, queued waiters ahead of the latency-critical thread have not acquired the lock yet, so we can bypass these waiters and let the latency-critical thread proceed immediately after the current lock holders release the lock.

\subsubsection{Identification of Blocking Locks} \label{lockPinpoint}

Before identifying blocking threads, we first need to identify the contended locks on which the latency-critical thread blocks.
Doing so is difficult on Android because many lock types and vendor-specific call paths make per-lock instrumentation impractical.
We take advantage of an interesting observation: long priority-inversion blockings are dominated by \emph{sleepable} locks. We refer to locks that put contending threads to sleep (e.g., \texttt{rwsem}) as sleepable locks, which makes their blockers sleep, later wake up, and may incur long runqueue delays before running again. 
In contrast, spinlocks busy-wait and do not trigger this sleep--wake--runqueue pattern. 
Our profiling results show that across 16 apps, spinlock contention has a P99.99 acquisition latency of $660 \pm 282\,\text{\textmu s}$, while contended \texttt{rwsem} acquisitions reach up to $210$\,ms (Figure~\ref{fig:motiv_tail}).

Guided by this observation, We take advantage of a common property of sleepable locks that under contention, a blocking thread eventually enters the scheduler’s shared sleep path. 
Thus, \ourSolution records a lightweight ``sleep fingerprint'' at the single scheduler sleep point and aggregates these fingerprints offline to identify the lock APIs most responsible for blockings on latency-critical threads.


These offline results remain valid as long as the kernel and vendor lock implementations and their sleep paths remain unchanged. The same contended lock operations continue to pass through the same scheduler sleep point and produce the same fingerprints.
Offline lock-identification analysis needs to be be re-run upon major kernel or vendor updates that may change lock implementations or call paths.
This approach eliminates per-lock instrumentation and reduces profiling effort. 
It is portable because it relies only on the scheduler’s shared sleep path, which is common across many systems.

\noindent\textbf{Implementation.}
We implement a bounded per-lock holder list to record the lock holders. 
To allow the latency-critical thread bypass queued waiters, we place it at the head of the lock’s waiter queue, ahead of ordinary waiters. 
On the Android kernel, the scheduler’s shared sleep path goes through \texttt{\_\_schedule()}. 
Thus, we hook \texttt{\_\_schedule()} to identify the locks blocking latency-critical threads, specifically right before a sleeping thread is switched out (i.e., just before \texttt{context\_switch()}).
We grab a short ``fingerprint'' of \emph{why} the thread is going to sleep: we record one stack frame at a chosen depth (\texttt{stackNum}) using \texttt{my\_dump\_stack(sym, stackNum)} and write the corresponding kernel symbol (\texttt{sym}) into the trace log.
Each sleep therefore produces a stable kernel symbol \texttt{sym} that points to the blocking primitive on the sleep path (e.g., \texttt{down\_write()} for a contended \texttt{rwsem} write).
We then aggregate these symbols offline to identify the lock APIs that dominate latency-critical threads' blocking. 

\subsection{Generality and Limitations} \label{generalization}

\noindent\textbf{Generality.}
\ourSolution is applicable to priority inversions on different latency-critical threads (e.g., UI thread, input dispatch, sensor processing, camera pipeline control, and XR/AR tracking).
\ourSolution is also not tied to a specific scheduler. Its core requirement is a \fastPathNoun that promptly runs labeled blocking threads by adjusting the scheduler's native ordering key. For example, on Android CFS, it lowers a labeled thread's \emph{vruntime}; on Linux EEVDF~\cite{linux_eevdf_docs}, it can similarly adjust a labeled thread's eligibility time and virtual deadline so that the thread is selected to run on CPU earlier.

\noindent\textbf{Limitations.}
\ourSolution shortens only scheduler-induced delays (i.e., runnable delays). It does not reduce the blocking thread’s critical-section latency and the time spent on sleeping (e.g., sleep for I/O).
We currently scope to a single latency-critical thread, and arbitration across multiple concurrent latency-critical threads is left to future work.

\section{Evaluation Setup} \label{sec:evaluateSetup}

In this section, we present the experimental platform, comparison mechanisms, workloads, and evaluation metrics.

\subsection{Experimental Platform} 
Our main experimental platform is a commercial smartphone, Google Pixel~8, with 8\,GB DRAM, 128\,GB flash storage, and 9 CPU cores (1$\times$3.0\,GHz Cortex-X3, 4$\times$2.45GHz Cortex-A715, 4$\times$2.15\,GHz Cortex-A510). The device runs Android~14 with Linux~5.15, where CFS is the default scheduler. 
Following previous reviewers’ comments, we use a high-end smartphone with a relatively new Android version to show that the priority-inversion problem remains prevalent and severe on modern mobile devices.
We also measure priority inversions on a Pixel~5 to show that the problem is not unique to Pixel~8, but is general across mobile devices. 
In this paper, we present the main results on Pixel~8, and include the Pixel~5 results in the supplementary materials.

\subsection{Comparison Mechanisms}  

We extensively compare our solution to the default kernel on Android, PI, RT-UI, and PE (state-of-the-art):

\noindent\textbf{Baseline:} We use the default Android kernel without modification, which runs Linux~5.15 kernel. The task scheduling scheme is widely used in modern mobile devices.

\noindent\textbf{Priority Inheritance (PI):} To reduce penalty, we implement and optimize priority inheritance for \texttt{rwsem}\footnote{UI threads are usually blocked on this lock. We take UI threads as a case study in our experiments.} by boosting the priority of the low-priority writer holding the lock when only higher-priority latency-critical thread is blocked. Thus, we eliminate the overhead associated with the global adoption of traditional priority inheritance. The boost is applied by adjusting the holder’s scheduling priority within CFS (i.e., prio value), without changing the scheduler class. The boost is revoked when the lock is released. 

    \noindent\textbf{Real-time UI (RT-UI):} We implement an intuitive solution that promotes the foreground app’s UI thread to a real-time scheduling class while it is in the foreground and restores its priority when it goes to the background. No changes are made to lock holders or synchronization.
    
   \noindent\textbf{Proxy Execution (PE):}
    \footnote{Due to porting constraints, PE is evaluated on an x86 Cuttlefish Android virtual device~\cite{corbet_lwn_proxy_exec_2023efish} configured to emulate Pixel~8’s 9-core CPU. In \S~\ref{cuttlefish_results}, Proxy Execution and SuperPass are both evaluated on Cuttlefish to ensure a fair comparison.}
   For fair comparison, we adapt PE to \texttt{rwsem} based on the original mutex-oriented design~\cite{pe_github}. When a latency-critical thread blocks on a \texttt{rwsem}, PE randomly chooses a lock holder to run instead of itself. PE handles one-to-one blocking but does not accelerate multiple concurrent readers in shared-lock scenarios.
   
    \noindent\textbf{\ourSolution:} Our solution \ourSolution is also implemented on \texttt{rwsem}. First, it makes the identified blocking threads run on CPU instantly via a scheduler \fastPathNoun. Thus, its performance outperforms the above mechanism. Second, it identifies most blocking threads (e.g., concurrent lock holders) of latency-critical threads with low overhead via a bounded list of lock holders.

In summary, PI and RT-UI capture two naive approaches: boosting the lock holder or prioritizing the latency-critical thread. PE represents the closest fair-scheduler-aware mechanism that accelerates blocking threads. Other approaches, such as PCP or reservation-based scheduling methods (e.g., SCHED\_DEADLINE), require different system assumptions (e.g., static task sets or explicit runtime/deadline assignments) and are therefore discussed in §8 rather than used as evaluation baselines.

\subsection{Workloads and Evaluation Metrics}  

\noindent\textbf{Workloads.} We use 16 apps across diverse categories (Table~\ref{tab:16_APPs}) as foreground applications. We exclude games, as their rendering workloads bypass the UI thread~\cite{game_rendering_one,game_rendering_three,UIjank1}, which we use as a case study of latency-critical threads in our evaluation. To avoid human bias in testing, we use automated scripts to interact with these foreground apps via adb shell input commands, ensuring reproducible user interactions.
In each test, we run 12 apps concurrently in the background to approximate realistic user behavior~\cite{acclaim, buildfire_app_statistics}. 
All experimental results in Sections \S~\ref{motivation}, \S~\ref{insights}, and \S~\ref{sec:improvement} are obtained using these realistic workloads.

\begin{table}[h!]
\centering
\vspace{-0.7em}
\small 
\setstretch{0.9} 
\caption{Sixteen Tested Applications}
\vspace*{-12pt}
\label{tab:16_APPs}
\begin{threeparttable} 
\begin{tabular}{@{}>{\bfseries}l>{\RaggedRight}p{0.5\linewidth}@{}}
\toprule
\textbf{Application Category} & \textbf{Applications} \\
\midrule
Social Media & 
Facebook, X, LinkedIn, Threads \\
\addlinespace[1pt]

Short Video & 
Tiktok, Instagram \\
\addlinespace[1pt]

Online Shopping & 
Amazon, Shein, Weee, Taobao \\
\addlinespace[1pt]

Browsing \& News & 
Firefox, Inshorts, WaPost \\
\addlinespace[1pt]

Booking \& Lifestyle & 
Booking, Ctrip, Alipay \\

\bottomrule
\end{tabular}
\begin{tablenotes}
\footnotesize
\item[1] WaPost represents Washington Post.
\end{tablenotes}
\end{threeparttable}
\vspace{-0.6em}
\end{table}

\noindent\textbf{Evaluation Metrics.}  
We measure a latency-critical thread's priority-inversion blocking duration directly in the kernel by recording two timestamps around the lock acquisition. When a latency-critical thread attempts to acquire a lock and finds it held by a lower-priority thread, we take a start timestamp using \texttt{ktime\_get()}. When the latency-critical thread acquires the lock successfully, we take an end timestamp and compute the elapsed time as the priority-inversion blocking duration, and report it via a customized \texttt{ftrace} tracepoint.
We report three complementary metrics:  
\begin{itemize}[leftmargin=*]
\vspace{-0.5em}
  \item {P99.9 Blocking Duration (ms)}: the 99.9th percentile of blocking durations of priority inversions on UI thread, capturing the key scenarios that are affected by priority inversion;
 \vspace{-0.5em} 
  \item {Number of Priority Inversion Blockings ($>$100\,\textmu s)}\footnote{The 100 \textmu s is used as a filtering threshold to exclude trivial events and focus on non-negligible priority inversions. These blockings matter because they can accumulate and, when overlapping the frame critical path, cause janky frames as shown in \S~\ref{longUIblock}. }: the number of priority inversion blockings exceeding 100\,\textmu s per minute, indicating how often noticeable delays occur;
\vspace{-0.5em}

  \item {Number of Janky Frames}: the number of dropped frames caused by priority inversions on UI thread, collected by Android \texttt{dumpsys} \texttt{framestats}~\cite{dumpsys}, directly reflecting user-perceived UI smoothness. 

\end{itemize}
Each foreground app was tested in ten one-minute runs. We report \emph{P99.9 Blocking Duration} over all runs. For \emph{Number of Priority Inversion blocking ($>$100\,\textmu s)} and \emph{Number of Janky Frames}, we report the average across runs with error bars showing minimum and maximum values.  

\section{Evaluation Results} \label{evaluation}

In this section, we comprehensively evaluate the effectiveness of \ourSolution, using UI threads as a case study and reduced janky frames as the user-experience metric. \S~\ref{sec:improvement} presents the main results, including reductions in priority-inversion counts and durations, improvements in janky frames, and the fairness impact on non-blocking threads. \S~\ref{sec:heavyWorkloads} evaluates robustness under heavier background workloads. \S~\ref{cuttlefish_results} compares \ourSolution with Proxy Execution (PE).

\subsection{Performance Improvement} \label{sec:improvement}
To evaluate the effectiveness of \ourSolution, we use three metrics. \S~\ref{sec:low_level_metric} reports the overall performance on reducing the number and duration of priority-inversion blockings on the latency-critical thread (UI thread in our evaluation).
\S~\ref{sec:high_level_metric} measures its impact on user experience, 
and \S~\ref{sec:starvation_safety} then verifies fairness impact on non-blocking threads.

\begin{figure}[h!]
\vspace{-0.9em}
  \centering
    \begin{subfigure}[t]{0.47\textwidth}
    \centering
    \includegraphics[width=\textwidth]{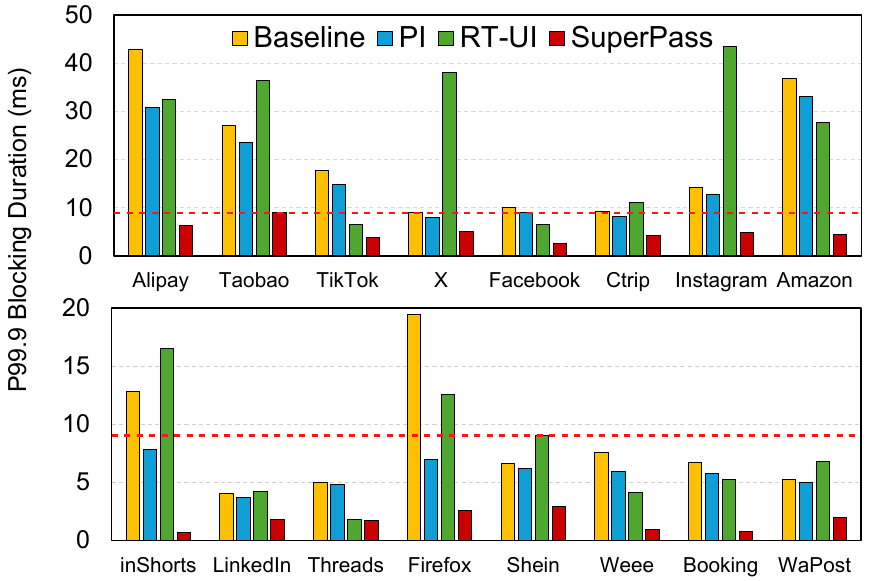}
    \caption{The 99.9th percentile durations of UI-thread priority-inversion blocking. Bars above the red line in 8.3 ms indicate risk of causing dropped frames.}
    \label{fig:evaluation_P999}
  \end{subfigure}
  
  \vspace{0.6em} 

  \begin{subfigure}[h!]{0.47\textwidth}
    \centering
    \includegraphics[width=\textwidth]{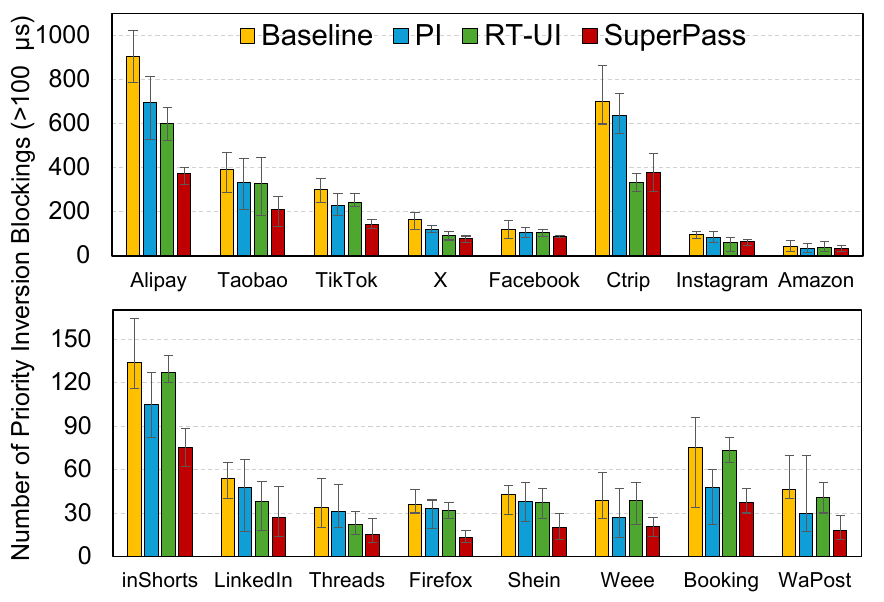}
    \caption{The number of UI-thread priority inversions ($>$ 100\,\textmu s).}
    \label{fig:evaluation_100us}
  \end{subfigure}
      
  \vspace{-1em}
  \caption{Overall performance (UI-thread instantiation). The duration and number of UI-thread priority-inversion blockings on Baseline, PI, RT-UI, and \ourSolution.
  }
  \vspace{-2em}
  \label{fig:overall_performance}
\end{figure}

\subsubsection{Overall Performance} 
\label{sec:low_level_metric}

Figures~\ref{fig:evaluation_100us} and \ref{fig:evaluation_P999} compare \ourSolution with Baseline, PI, and RT-UI under a realistic workload (one foreground app with 12 background apps). 
On average, \ourSolution achieves a 72.0\% reduction on P99.9 blocking duration across 16 apps compared to Baseline and delivers a $4.6\times$ improvement over PI. 
RT-UI increases tail latencies in some cases because raising UI thread to real-time priority reduces CPU time for CFS-scheduled blocking threads of UI thread.
For the number of UI‑thread priority inversion blocking, \ourSolution achieves a 47.7\% reduction relative to Baseline, corresponding to $2.2\times$ fewer blocks than PI and $2.5\times$ fewer than RT-UI. \ourSolution reduces the duration longer than $>$ 100\,\textmu s by effectively accelerating blocking threads of UI thread.
In conclusion, these results demonstrate that \ourSolution more effectively mitigates both the number and duration of priority-inversion blockings on UI thread.

\begin{figure}[h!]
\vspace{-0.8em}
\centering
\includegraphics[width=0.43\textwidth]{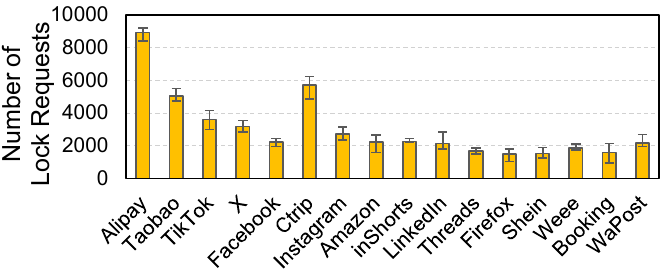}
\vspace{-0.5em}
\caption{
Number of lock requests issued by the foreground app’s UI thread (each test runs for one minute).}
 \vspace{-0.5em}
\label{fig:lock_request_count}
\end{figure}

\noindent\textbf{Variation Across Apps.}
Figures~\ref{fig:overall_performance} show that the severity of priority inversion on UI thread varies widely across apps. Figure~\ref{fig:lock_request_count} suggests that this depends on lock-acquisition intensity. Apps like Alipay produce thousands of lock requests and thus face more priority inversions, while Firefox acquires fewer locks and correspondingly has fewer priority inversions. 
These results indicate that lock-acquisition intensity is a key factor in an app’s susceptibility to severe priority-inversion blockings on UI thread.

\noindent\textbf{Performance Breakdown.} 
As described in \S~\ref{blockerDetector}, a contended lock can delay the latency-critical thread through three kinds of lock participants: (i) a writer holder that currently owns the lock, (ii) queued waiters ahead of the requester, and (iii) reader holders that collectively hold a shared lock.
\ourSolution handles them with three corresponding mechanisms: 
\textit{Writers Acceleration} labels and expedites a writer holder via the scheduler \fastPathNoun; 
\textit{Waiters Bypass} places the blocked latency-critical thread at the head of the lock’s wait queue;
\textit{Readers Acceleration} does the same for a bounded set of reader holders.

\begin{figure}[h!]
\vspace{-0.5em}
  \centering
  \begin{subfigure}[t]{0.46\textwidth}
    \centering
    \includegraphics[width=\textwidth]{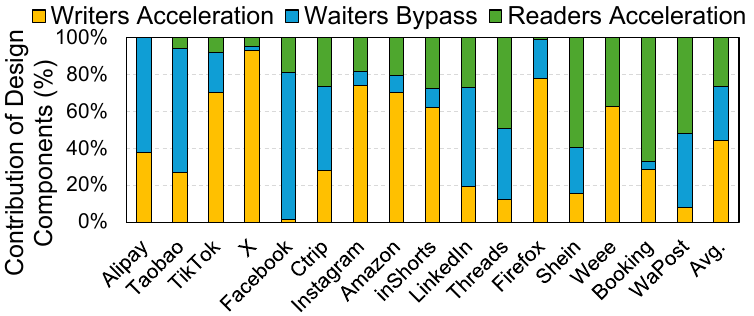}
    \caption{
    The P99.9 priority-inversion blockings on UI thread.}
    \label{fig:ratio_tail}
  \end{subfigure}
  
  \vspace{0.3em} 
  
  \begin{subfigure}[t]{0.46\textwidth}
    \centering
    \includegraphics[width=\textwidth]{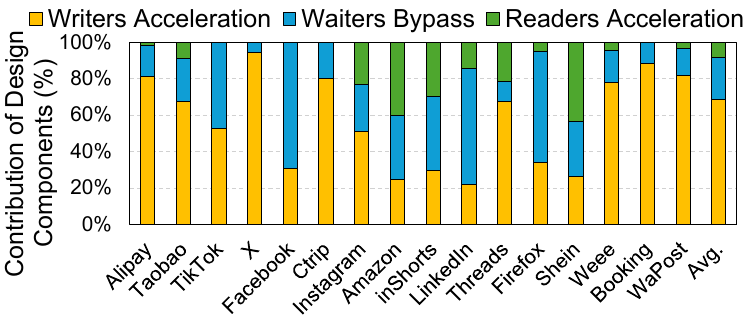}
    \caption{The number of priority-inversion blockings exceeding 100\,\textmu s on UI thread.}
    \label{fig:ratio_count}
  \end{subfigure}
  
  \vspace{-0.5em}
  
  \caption{Performance breakdown.}
  \label{fig:PB_ratio}
  \vspace{-1em}
\end{figure}

We quantify the contribution of these three mechanisms by enabling them incrementally.
Figure~\ref{fig:PB_ratio} breaks down the resulting improvement. For each app, we normalize the total improvement over the baseline to 100\%, and the stacked bars show the fraction of this total attributable to the three components.
In Figure~\ref{fig:ratio_tail}, averaged across apps, \textit{Writers Acceleration}, \textit{Waiters Bypass}, and \textit{Readers Acceleration} contribute 44.4\%, 29.2\%, and 26.4\% of the total reduction in the P99.9 priority-inversion blocking duration on UI thread, respectively.
In Figure~\ref{fig:ratio_count}, the corresponding shares for reducing the number of priority-inversion blocking events ($>100\,\text{\textmu s}$) are 58.3\%, 31.2\%, and 10.4\%.
Contributions vary across apps because their lock-contention patterns and read/write mixes differ. 

\subsubsection{User Experience Impact: Janky Frame} 
\label{sec:high_level_metric}
\begin{figure}[h!]
\centering
\includegraphics[width=0.46\textwidth]{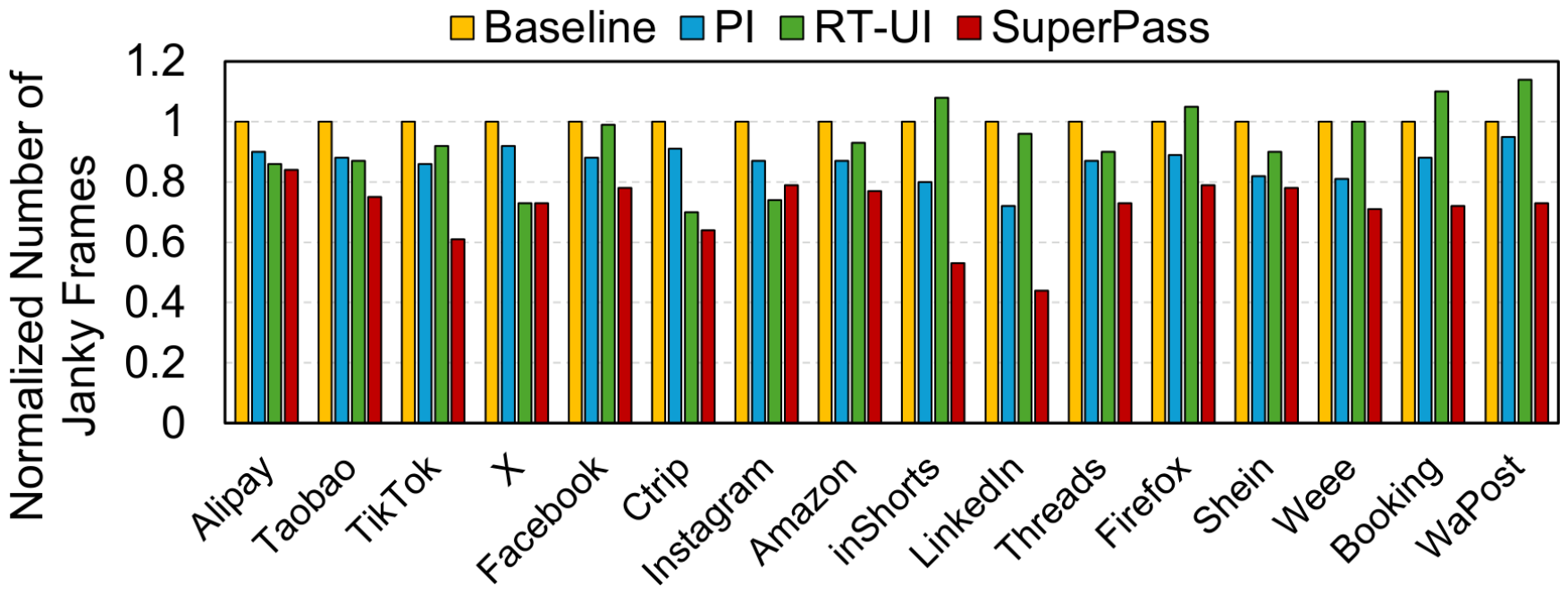}
\vspace{-0.9em}
\caption{Number of janky frames (normalized to Baseline).}
\label{fig:evaluation_jankyframes}
\end{figure}

To evaluate \ourSolution's user-experience impact, we measured janky frames under the same setup as in \S~\ref{sec:low_level_metric}. 
Figure~\ref{fig:evaluation_jankyframes} shows that \ourSolution reduces the number of janky frames by 29.2\% compared with the Baseline on average, which corresponds to $2.1\times$ fewer janky frames than PI and $4.1\times$ fewer than RT-UI. 
RT-UI even increased the janky frames for some apps, as it may block other important system real-time threads. 
In conclusion, \ourSolution consistently outperforms PI and RT-UI across all 16 apps in reducing janky frames.

\subsubsection{Faireness Impact on Non-Blocking Threads}\label{sec:starvation_safety}
\begin{figure}[h!]
\vspace{-0.5em}
  \centering
  \includegraphics[width=0.48\textwidth]{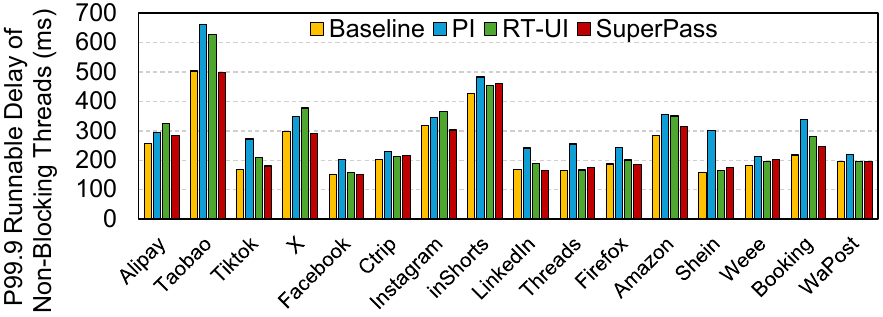}
  \vspace{-1.8em}
  \caption{P99.9 runnable delay of low-priority threads not blocking UI thread across 16 apps.}
  \vspace{-1em}
  \label{fig:starvation_p999_runnable}
\end{figure}

To evaluate the faireness impact of \ourSolution on non-blocking threads, we measure their runnable delays under the same realistic setup as in \S~\ref{sec:low_level_metric}.
Figure~\ref{fig:starvation_p999_runnable} reports the P99.9 runnable delays of low-priority non-blocking threads across 16 apps. 
PI increases the P99.9 runnable delay by 32.6\% on average, indicating longer blocking for threads not blocking UI thread. RT-UI also raises this metric by 13.7\% on average. 
In contrast, \ourSolution increases this metric by only 4.7\% on average, resulting in nearly baseline-level overhead across all apps.
In conclusion, these results show that \ourSolution have small faireness impact on non-blocking threads, consistent with the safeguards in \S~\ref{sec:safeguards}.
Although strict fairness cannot be guaranteed, \ourSolution still outperforms existing methods by respecting smartphone workloads where foreground apps and UI threads are prioritized for user experience.

\subsection{Performance under Heavy Workloads} \label{sec:heavyWorkloads}

To emulate heavy usage, we add a synthetic CPU workload with busy-loop workers pinned to cores. We use this workload only in this subsection to evaluate \ourSolution\ under heavier background contention and to verify both its robustness and fairness impact on non-blocking threads.

\begin{figure}[h!]
\vspace{-0.6em}
  \centering
  \includegraphics[width=0.48\textwidth]{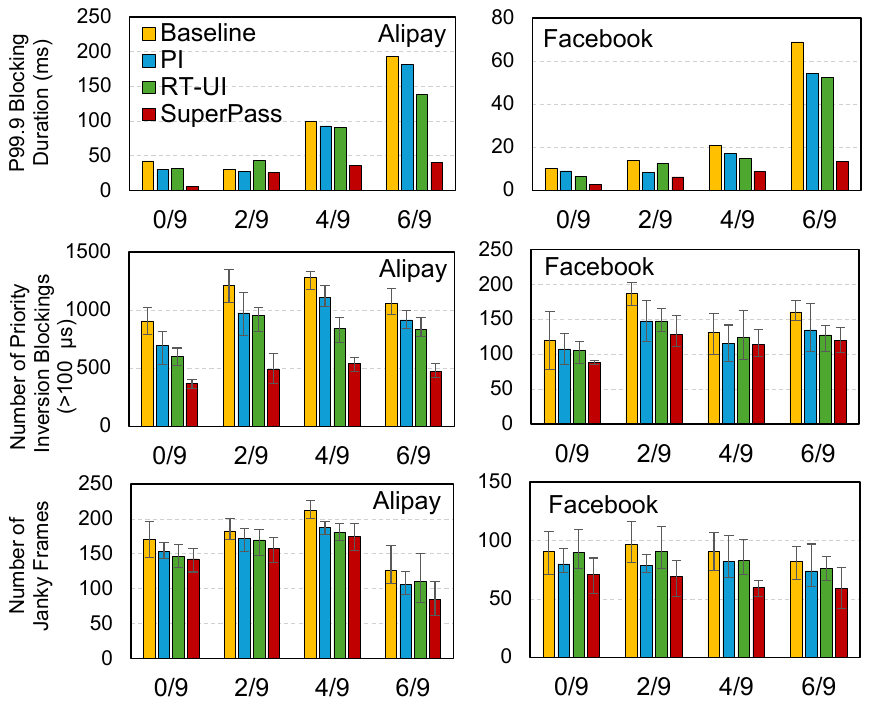}
 \vspace{-2em} 
  \caption{Performance under heavy workload. Each row reports, from top to bottom, the P99.9 blocking duration, the number of blocks >100\,\textmu s, and the number of janky frames for Alipay and Facebook. The x-axis “$k/9$” denotes that $k$ of the 9 CPU cores are preoccupied by the busy-loop workers.}
  \vspace{-0.6em}
  \label{fig:sensitivity_all}
\end{figure}

Specifically, We pin $k\in\{2,4,6\}$ busy-loop workers to distinct cores while running 12 background apps concurrently in the background. 
All busy-loop workers are normal threads scheduled by CFS. Figure~\ref{fig:sensitivity_all} reports two representatives: Alipay (many lock requests, frequent inversions) and Facebook (fewer lock requests). We report all results of 16 apps in the supplementary file.
In conclusion, across all workloads, \ourSolution consistently reduces the number and duration of priority-inversion blockings on UI thread, and correspondingly lowers the number of janky frames, relative to Baseline, PI, and RT-UI, demonstrating robustness under heavy background contention.

\begin{figure}[h!]
\centering
\vspace{-0.6em}
\includegraphics[width=0.48\textwidth]{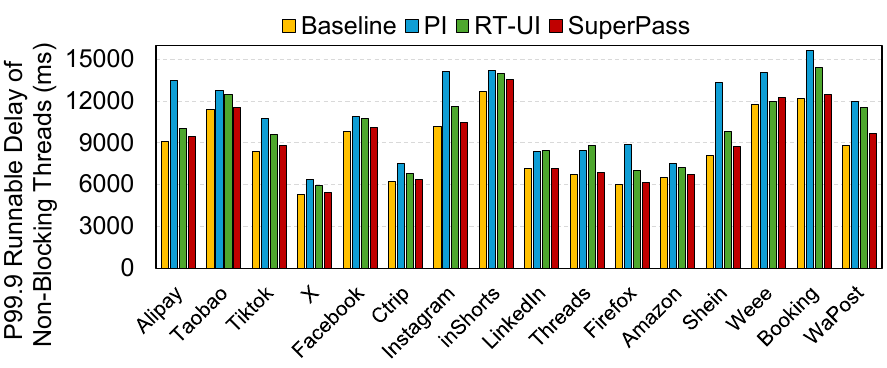}
\vspace{-1.9em}
\caption{P99.9 runnable delay of low-priority threads not blocking UI thread across 16 apps, under heavy background workload with 6 busy-loop workers.}
\label{fig:G1234_12bak_75_penalty}
\vspace{-1.4em}
\end{figure}

We also assess fairness impact on non-blocking threads using the P99.9 runnable delay under heavy background workload with \(k=6/9\) cores occupied. Figure~\ref{fig:G1234_12bak_75_penalty} shows that \ourSolution remains close to Baseline across apps (3.6\% increase), while PI and RT-UI raise the runnable delay of non-blocking threads by 27.7\% and 14.5\%, respectively. 
In conclusion, the results confirms that our safeguards limit the impact of \ourSolution on the fairness of non-blocking threads, even under severe CPU contention.

\subsection{Comparison with Proxy Execution} \label{cuttlefish_results}

We compare \ourSolution against Proxy Execution (PE)~\cite{lwn_proxy_exec_eurosys_comments,corbet_lwn_proxy_exec_2023} on Cuttlefish (9 vCPUs, 8\,GB RAM) over five foreground apps with 12 background apps.
PE is difficult to deploy directly on Pixel~8 due to porting constraints, as it was originally designed and implemented for Linux kernel mutexes rather than Android’s \texttt{rwsem}. 

\begin{figure}[h!]
\vspace{-0.3em}
\centering
\includegraphics[width=0.48\textwidth]{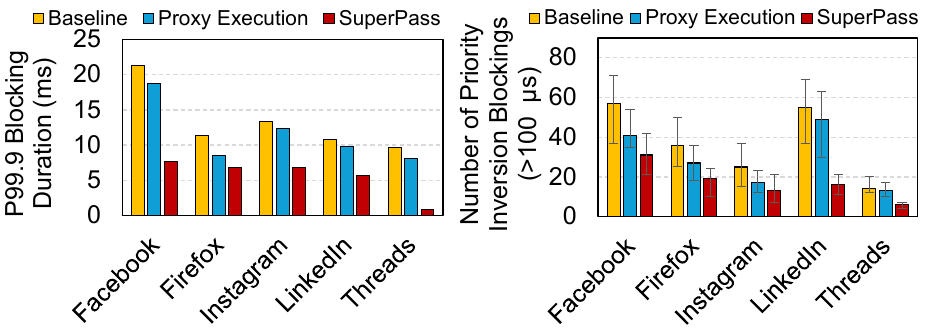}
\vspace{-2em}
\caption{Comparison with Proxy Execution on Cuttlefish.}
\vspace{-1em}
\label{fig:evaluation_cuttlefish}
\end{figure}

Figure~\ref{fig:evaluation_cuttlefish} shows that PE reduces the number of priority inversions on UI thread with blocking duration longer than 100\,\textmu s by 20.6\% and the P99.9 blocking duration by 14.0\%. 
In comparison, \ourSolution achieves reductions of 53.7\% and 51.0\% (i.e., $2.61\times$ and $3.64\times$ larger improvements than PE, respectively).
This advantage comes from two factors: 
(i) broader coverage of blocking threads, since \ourSolution accelerates multiple blocking threads such as \texttt{rwsem} readers that PE does not cover;
and 
(ii) faster acceleration, because once a blocking thread is labeled, \ourSolution grants it CPU immediately via the scheduler \fastPathNoun, whereas PE runs a lock holder only when the blocked high-priority waiter is scheduled, leaving additional runqueue delay.
On Cuttlefish, we observe fewer and shorter priority-inversion blockings on the UI-thread for three reasons. (i) it lacks big.LITTLE  DVFS/thermal limits used in physical smartphones~\cite{ARM_bigLITTLE,AOSP_Thermal}, (ii) it omits OEM/background services and sensor stacks~\cite{corbet_lwn_proxy_exec_2023efish}, and (iii) it uses a virtual display pipeline~\cite{LWN_display}. 
In conclusion, \ourSolution outperforms PE in reducing both the frequency and duration of priority-inversion blockings on UI thread.

\subsection{Overhead Analysis}  
\noindent\textbf{Memory Overhead.} We implement \ourSolution in the Android kernel with minimal modifications. We add a per-thread flag in the struct \texttt{task\_struct} to mark the blocking threads of the latency-critical threads (1\,B) and a list in each 
rwsem to track up to eight reader holders (\(8\times\mathrm{sizeof}(\texttt{void*})=64\)B). Writers reuse the existing \texttt{owner} field, so no extra space is added for writer holders. 

\noindent\textbf{CPU Overhead.} At runtime, the lock-level detector for finding blocking threads works only when the latency-critical threads fail to obtain a lock. The detector performs \(O(1)\) checks (latency-critical thread/label status and lock holders' priority). On the lock side, each lock holder is inserted into (and removed from) the end of the bounded list in \(O(1)\). Overall, the measured system-wide CPU overhead is \(\approx 0.74\%\) (details in the supplementary file).

\section{Related Work}

To the best of our knowledge, there is no previous work targeting priority inversions of latency-critical threads on Android.
The existing solutions designed for real-time systems fail to address the problem on Android and may introduce high overheads.
To address this gap, we propose a new solution for Android, \ourSolution, which is inspired by two insights: first, the long priority inversion blockings are driven by the accumulated CPU waiting time of low-priority blocking threads on the scheduler runqueue. Second, on Android, even latency-critical threads could be blocked by many concurrent readers, tracking a limited number of readers suffices to achieve good responsiveness while maintaining low overhead in most cases.
We conduct comprehensive comparisons with both the default solution in modern smartphones and the state-of-the-art solutions. The remainder of this section reviews related work on priority inversion mitigation and lock-optimization-based potential solutions.

\noindent\textbf{Priority Inversion Mitigation.}
Typical solutions in real-time systems for priority inversion include Priority Inheritance (PI) and Priority Ceiling Protocol 
(PCP)~\cite{PIP_PCP_1990,goodenough1988priority,cheng2007implementation}. 
The core idea of PI is temporarily raising a low-priority lock holder to the blocked thread’s priority until the lock is released and propagating this boost transitively across chains of blocking threads.
We implement this core idea on \texttt{rwsem} and compare it with our solution. 
One of PI's variants is PI for priority-ordered spin locks~\cite{PIP2}.
PI-based solutions share a common limitation that they assume priority raising leads to immediate CPU access, which does not hold under Android’s CFS.
The core idea of PCP is to assign each lock a priority ceiling equal to the highest priority of any task that may use it and to execute a task at this ceiling while it holds the lock. 
PCP has several variants, including the Stack Resource Policy (SRP)~\cite{PCP2}, the immediate priority ceiling protocol~\cite{linux-not-important}, and multiprocessor extensions~\cite{multiprocessor2SharedMemoryMultiprocessors,multiprocessor4RealTimeTasks}. PCP-based solutions require prior knowledge of all potential lock users and their priorities, which is impractical in dynamic Android workloads.

Recently, Proxy Execution~\cite{lwn_proxy_exec_eurosys_comments,corbet_lwn_proxy_exec_2023,stultz_lkml_proxy_exec_2023,pe_github} has been proposed as a mechanism to accelerate lock holders by executing them on behalf of blocked threads. We evaluate it as a comparison mechanism. Its general form is allocation inheritance~\cite{allocation_inheritance}, which generalizes priority inheritance by transferring processor allocation or execution rights to the lock holder. These approaches also aim to unblock waiters. \ourSolution differs by targeting Android shared-lock multi-reader blocking and enabling immediate execution of blocking threads, rather than waiting for the high-priority thread’s allocated CPU time.

\noindent\textbf{Lock-Optimization-Based Potential Solutions.}
Reader-writer lock designs have also been optimized to reduce contention, improve fairness, or lower tail latency. 
Examples include phase-fair reader-writer locks~\cite{phase_fair_rwlock}, read-biased designs such as BRAVO~\cite{bravo_rwlock}, and queue-management or shuffling-based designs that optimize fairness and scalability~\cite{shfl_lock,hoshino2025_rwlock}. 
Read-optimized alternatives such as RCU further reduce read-side blocking by changing the synchronization model~\cite{McKenney2020RCU18,Kim2022PerformanceAO,Corbet2024RCUAPI}. 
These approaches are complementary to \ourSolution. 
They redesign the lock or synchronization primitive itself, whereas \ourSolution targets scheduler-induced runnable delays on the blocking dependency chain in existing Android shared locks. 
This distinction is important because Figure~\ref{fig:critical_section_vs_runnable} shows that long priority-inversion blockings are dominated by the runnable delay of blocking threads rather than lock critical-section time. 
Therefore, lock redesign alone does not directly address the main bottleneck we target, while \ourSolution can coexist with improved lock implementations.

\section{Conclusion}
In this work, we investigate the priority inversion of latency-critical threads in Android systems. Using the foreground UI thread as a representative case study,
we find that it frequently causes prolonged blocking durations and dropped frames that degrade the user experience of mobile devices.
Existing solutions designed for real-time systems are ineffective and could introduce fairness issues. 
To address this problem, we proposed \ourSolution, a lightweight kernel mechanism that effectively reduces the blockings of the latency-critical threads with safeguards to limit impact on fairness. 
We evaluate \ourSolution on a Google Pixel~8. The results show that \ourSolution significantly reduces the priority-inversion blocking durations compared to both the default scheduler and state-of-the-art approaches.




\bibliographystyle{ACM-Reference-Format}
\bibliography{main}

@inproceedings{liu2020removing,
  title={On removing algorithmic priority inversion from mission-critical machine inference pipelines},
  author={Liu, Shengzhong and Yao, Shuochao and Fu, Xinzhe and Tabish, Rohan and Yu, Simon and Bansal, Ayoosh and Yun, Heechul and Sha, Lui and Abdelzaher, Tarek},
  booktitle={2020 IEEE Real-Time Systems Symposium (RTSS)},
  pages={319--332},
  year={2020},
  organization={IEEE}
}

@inproceedings {acclaim,
author = {Yu Liang and Jinheng Li and Rachata Ausavarungnirun and Riwei Pan and Liang Shi and Tei-Wei Kuo and Chun Jason Xue},
title = {Acclaim: Adaptive Memory Reclaim to Improve User Experience in Android Systems},
booktitle = {2020 USENIX Annual Technical Conference (USENIX ATC 20)},
year = {2020},
isbn = {978-1-939133-14-4},
pages = {897--910},
publisher = {USENIX Association},
month = jul
}

@inproceedings{CFSone,
author = {Justinien Bouron and Sebastien Chevalley and Baptiste Lepers and Willy Zwaenepoel and Redha Gouicem and Julia Lawall and Gilles Muller and Julien Sopena},
title = {The Battle of the Schedulers: {FreeBSD} {ULE} vs. Linux {CFS}},
booktitle = {2018 USENIX Annual Technical Conference (USENIX ATC 18)},
year = {2018},
isbn = {978-1-931971-44-7},
address = {Boston, MA},
pages = {85--96},
publisher = {USENIX Association},
month = jul
}

@inproceedings{CFStwo,
author = {Lozi, Jean-Pierre and Lepers, Baptiste and Funston, Justin and Gaud, Fabien and Qu\'{e}ma, Vivien and Fedorova, Alexandra},
title = {The Linux scheduler: a decade of wasted cores},
year = {2016},
isbn = {9781450342407},
publisher = {Association for Computing Machinery},
address = {New York, NY, USA},
booktitle = {Proceedings of the Eleventh European Conference on Computer Systems},
articleno = {1},
numpages = {16},
location = {London, United Kingdom},
series = {EuroSys '16}
}

@ARTICLE{PIP_PCP_1990,
  author={Sha, L. and Rajkumar, R. and Lehoczky, J.P.},
  journal={IEEE Transactions on Computers}, 
  title={Priority inheritance protocols: an approach to real-time synchronization}, 
  year={1990},
  volume={39},
  number={9},
  pages={1175-1185},
}

@INPROCEEDINGS{PCP2,
  author={Baker, T.P.},
  booktitle={1990 Proceedings 11th Real-Time Systems Symposium}, 
  title={A stack-based resource allocation policy for realtime processes}, 
  year={1990},
  volume={},
  number={},
  pages={191-200},
}

@INPROCEEDINGS{PIP2,
  author={Cai-Dong Wang and Takada, H. and Sakamura, K.},
  booktitle={Proceedings Second International Symposium on Parallel Architectures, Algorithms, and Networks (I-SPAN'96)}, 
  title={Priority inheritance spin locks for multiprocessor real-time systems}, 
  year={1996},
  volume={},
  number={},
  pages={70-76},
}

@inproceedings{PI_concept1,
author = {Locke, D. and Sha, L. and Rajikumar, R. and Lehoczky, J. and Burns, G.},
title = {Priority inversion and its control: An experimental investigation},
year = {1988},
isbn = {0897912950},
publisher = {Association for Computing Machinery},
address = {New York, NY, USA},
booktitle = {Proceedings of the Second International Workshop on Real-Time Ada Issues},
pages = {39–42},
numpages = {4},
location = {Moretonhampstead, Devon, England},
series = {IRTAW '88}
}

@article{PI_concept2,
title = {Transaction scheduling protocols for controlling priority inversion: A review},
journal = {Computer Science Review},
volume = {35},
pages = {100215},
year = {2020},
issn = {1574-0137},
author = {Sarvesh Pandey and Udai Shanker},
}

@inproceedings{abeni1998cbs,
  author       = {Abeni, Luca and Buttazzo, Giorgio},
  title        = {Integrating Multimedia Applications in Hard Real-Time Systems},
  booktitle    = {Proceedings of the IEEE Real-Time Systems Symposium (RTSS)},
  year         = {1998}
}

@inproceedings{hierarchicalServers,
  author       = {Lipari, Giuseppe and Bini, Enrico},
  title        = {A Framework for Hierarchical Scheduling},
  booktitle    = {Proceedings of the IEEE Real-Time Systems Symposium (RTSS)},
  year         = {2003}
}

@techreport{priorityInversionFormation,
author = {Babaoglu, O. and Marzullo, K. and Schneider, F.},
title = {A Formalization of Priority Inversion},
year = {1993},
publisher = {University of Bologna},
}

@inproceedings{rt-important,
  title={Handling of Priority Inversion Problem in RT-Linux using Priority Ceiling Protocol},
  author={D. Silambarasan and M RamanathaVenkatesan},
  year={2016}
}

@article{linux-not-important,
author = {Carminati, Andreu and Oliveira, Rômulo and Luís, Fernando and Friedrich},
year = {2012},
month = {03},
pages = {},
title = {Implementation and Evaluation of the Synchronization Protocol Immediate Priority Ceiling in PREEMPT-RT Linux},
volume = {7},
journal = {Journal of Software},
}

@misc{pixel8wiki,
  title = {Google Pixel 8},
  howpublished = "\url{https://en.wikipedia.org/wiki/Pixel_8}",
  year = {2024}, 
  note = "[Online; accessed 20-March-2026]",
}

@misc{schedDeadline,
  title = {SCHED\_DEADLINE Linux Documentation},
  howpublished = "\url{https://www.kernel.org/doc/Documentation/scheduler/sched-deadline.txt}",
  year = {2024}, 
  note = "[Online; accessed 20-March-2026]",
}

@misc{UIjank1,
  title = {Slow rendering},
  howpublished = "\url{https://developer.android.com/topic/performance/vitals/render}",
  year = {2024}, 
  note = "[Online; accessed 20-March-2026]",
}

@misc{UIjank2,
  title = {UI jank detection},
  howpublished = "\url{https://developer.android.com/studio/profile/jank-detection}",
  year = {2024}, 
  note = "[Online; accessed 20-March-2026]",
}

@misc{buildfire_app_statistics,
  title = {Mobile App Usage \& Download Statistics},
  author       = {{Google}},
  howpublished = "\url{https://buildfire.com/app-statistics/}",
  year = {2024}, 
  note = "[Online; accessed 20-March-2026]",
}

@misc{linux_rwsem_owner_overhead,
  title        = {kernel/locking/rwsem.c (comment on overhead of tracking readers)},
  howpublished = {\url{https://git.zx2c4.com/wireguard-linux/tree/kernel/locking/rwsem.c}},
  note         = {Source comment: “Ideally we would like to track all the readers that own a rwsem, but the overhead is simply too big. [Online; accessed 20-March-2026]”},
  year         = {2023}
}

@misc{corbet_lwn_proxy_exec_2023efish,
  title = {Cuttlefish virtual Android devices},
  howpublished = "\url{https://source.android.com/docs/devices/cuttlefish}",
  year = {2025}, 
  note = "[Online; accessed 20-March-2026]",
}

@misc{pe_github,
  title = {Proxy Execution Github},
  howpublished = "\url{https://github.com/johnstultz-work/linux-dev/tree/proxy-exec-v6-6.6}",
  year = {2025}, 
  note = "[Online; accessed 20-March-2026]",
}

@misc{game_rendering_one,
  title = {Learn about rendering in game loops},
  howpublished = "\url{https://developer.android.com/games/develop/gameloops}",
  year = {2025}, 
  note = "[Online; accessed 20-March-2026]",
}

@INPROCEEDINGS{allocation_inheritance,
  author={Holman, P. and Anderson, J.H.},
  booktitle={Proceedings 14th Euromicro Conference on Real-Time Systems. Euromicro RTS 2002}, 
  title={Object sharing in Pfair-scheduled multiprocessor systems}, 
  year={2002},
  volume={},
  number={},
  pages={111-120}
}

@misc{linux_eevdf_docs,
  title        = {EEVDF Scheduler},
  howpublished = {The Linux Kernel Documentation},
  year         = {2026},
  url          = {https://docs.kernel.org/scheduler/sched-eevdf.html},
  note = "[Online; accessed 20-March-2026]",
}

@techreport{ARM_bigLITTLE,
  author       = {{ARM Ltd.}},
  title        = {{big.LITTLE Technology: The Future of Mobile}},
  institution  = {ARM Whitepaper},
  year         = 2013,
  url          = {https://developer.arm.com/-/media/Files/pdf/white-paper/big-little-technology-the-future-of-mobile.pdf}
}

@misc{AOSP_Thermal,
  title = {Thermal mitivation},
  howpublished = "\url{https://source.android.com/docs/core/power/thermal-mitigation}",
  year = {2025}, 
  note = "[Online; accessed 20-March-2026]",
}

@misc{ftrace,
  title = {ftrace - Function Tracer},
  howpublished = "\url{https://docs.kernel.org/trace/ftrace.html}",
  year = {2025}, 
  note = "[Online; accessed 20-March-2026]",
}

@misc{dumpsys,
  title = {Android dumpsys},
  howpublished = "\url{https://developer.android.com/tools/dumpsys}",
  year = {2025}, 
  note = "[Online; accessed 20-March-2026]",
}

@online{LWN_display,
  author       = {Alessio Balsini},
  title        = {{Scheduling for the Android display pipeline}},
  year         = 2020,
  howpublished = {LWN.net, Jan 16, 2020},
  url          = {https://lwn.net/Articles/809545/},
note = "[Online; accessed 20-March-2026]",
}

@article{McKenney2020RCU18,
author = {McKenney, Paul E. and Fernandes, Joel and Boyd-Wickizer, Silas and Walpole, Jonathan},
title = {RCU Usage In the Linux Kernel: Eighteen Years Later},
year = {2020},
issue_date = {July 2020},
publisher = {Association for Computing Machinery},
address = {New York, NY, USA},
volume = {54},
number = {1},
issn = {0163-5980},
month = aug,
pages = {47–63},
numpages = {17}
}

@article{Kim2022PerformanceAO,
  title={Performance Analysis of RCU-Style Non-Blocking Synchronization Mechanisms on a Manycore-Based Operating System},
  author={Chan-Kyung Kim and Eu-teum Choi and Mingyun Han and Seongjin Lee and Jaeho Kim},
  journal={Applied Sciences},
  year={2022},
}

@article{Corbet2024RCUAPI,
  author  = {Paul E. McKenney},
  title   = {The RCU API, 2024 edition},
  journal = {LWN.net},
  year    = {2024},
  month   = sep,
  url     = {https://lwn.net/Articles/988638/}
}

@misc{lwn_proxy_exec_eurosys_comments,
  author       = {Jonathan Corbet},
  title        = {What remains to be done for proxy execution},
  howpublished = {LWN.net},
  year         = {2023},
  month        = jun,
  url          = {https://lwn.net/Articles/953438/},
note = "[Online; accessed 20-March-2026]",
}

@misc{corbet_lwn_proxy_exec_2023,
  author       = {Jonathan Corbet},
  title        = {Addressing priority inversion with proxy exettion},
  howpublished = {LWN.net},
  year         = {2023},
  month        = jun,
  url          = {https://lwn.net/Articles/934114/},
note = "[Online; accessed 20-March-2026]",
}

@misc{stultz_lkml_proxy_exec_2023,
  author       = {John Stultz},
  title        = {[PATCH v3 00/14] Generalized Priority Inheritance and Proxy Execution},
  howpublished = {Linux Kernel Mailing List (LKML)},
  year         = {2023},
  month        = apr,
  url          = {https://lkml.org/lkml/2023/4/11/17},
note = "[Online; accessed 20-March-2026]",
}

@misc{game_rendering_three,
  title = {SurfaceView and GLSurfaceView},
  howpublished = "\url{https://source.android.com/docs/core/graphics/arch-sv-glsv}",
  year = {2025}, 
  note = "[Online; accessed 20-March-2026]",
}

@article{phase_fair_rwlock,
    author = {Brandenburg, Bj\"{o}rn B. and Anderson, James H.},
    title = {Spin-based reader-writer synchronization for multiprocessor real-time systems},
    year = {2010},
    issue_date = {September 2010},
    publisher = {Kluwer Academic Publishers},
    address = {USA},
    volume = {46},
    number = {1},
    issn = {0922-6443},
    doi = {10.1007/s11241-010-9097-2},
    month = sep,
    pages = {25–87},
    numpages = {63}
}

@inproceedings {bravo_rwlock,
    author = {Dave Dice and Alex Kogan},
    title = {{BRAVO{\textemdash}Biased} Locking for {Reader-Writer} Locks},
    booktitle = {2019 USENIX Annual Technical Conference (USENIX ATC 19)},
    year = {2019},
    isbn = {978-1-939133-03-8},
    address = {Renton, WA},
    pages = {315--328},
    publisher = {USENIX Association},
    month = jul
}

@inproceedings{shfl_lock,
    author = {Kashyap, Sanidhya and Calciu, Irina and Cheng, Xiaohe and Min, Changwoo and Kim, Taesoo},
    title = {Scalable and practical locking with shuffling},
    year = {2019},
    isbn = {9781450368735},
    publisher = {Association for Computing Machinery},
    address = {New York, NY, USA},
    doi = {10.1145/3341301.3359629},
    booktitle = {Proceedings of the 27th ACM Symposium on Operating Systems Principles},
    pages = {586–599},
    numpages = {14},
    location = {Huntsville, Ontario, Canada},
    series = {SOSP '19}
}

@inproceedings{hoshino2025_rwlock,
    author = {Hoshino, Takashi and Taura, Kenjiro},
    title = {Fairer and More Scalable Reader-Writer Locks by Optimizing Queue Management},
    year = {2025},
    isbn = {9798400714436},
    publisher = {Association for Computing Machinery},
    address = {New York, NY, USA},
    booktitle = {Proceedings of the 30th ACM SIGPLAN Annual Symposium on Principles and Practice of Parallel Programming},
    pages = {115–127},
    numpages = {13},
    location = {Las Vegas, NV, USA},
    series = {PPoPP '25}
}

@INPROCEEDINGS{multiprocessor2SharedMemoryMultiprocessors,
  author={Rajkumar, R.},
  booktitle={Proceedings.,10th International Conference on Distributed Computing Systems}, 
  title={Real-time synchronization protocols for shared memory multiprocessors}, 
  year={1990},
  volume={},
  number={},
  pages={116-123},
}

@article{multiprocessor4RealTimeTasks,
author = {Chen, Zewei and Lei, Hang and Yang, Maolin and Liao, Yong and Qiao, Lei},
title = {A Hierarchical Hybrid Locking Protocol for Parallel Real-Time Tasks},
year = {2021},
issue_date = {October 2021},
publisher = {Association for Computing Machinery},
address = {New York, NY, USA},
volume = {20},
number = {5s},
issn = {1539-9087},
journal = {ACM Trans. Embed. Comput. Syst.},
month = {sep},
articleno = {86},
numpages = {22}
}

@article{goodenough1988priority,
  title={The priority ceiling protocol: A method for minimizing the blocking of high priority Ada tasks},
  author={Goodenough, John B and Sha, Lui},
  journal={ACM SIGAda Ada Letters},
  volume={8},
  number={7},
  pages={20--31},
  year={1988},
  publisher={ACM New York, NY, USA}
}

@article{cheng2007implementation,
  title={The implementation of the priority ceiling protocol in Ada-2005},
  author={Cheng, Albert MK and Ras, James},
  journal={ACM SIGAda Ada Letters},
  volume={27},
  number={1},
  pages={24--39},
  year={2007},
  publisher={ACM New York, NY, USA}
}

\appendix

\end{document}